\documentclass{JHEP}

\usepackage{amssymb}

\usepackage{amsfonts}

\usepackage{amsbsy}

\usepackage{epsfig}

\newcommand{\im}{\mathop{\rm Im}\nolimits}
\newcommand{\re}{\mathop{\rm Re}\nolimits}

\renewcommand{\cal}{\mathcal}

\def\IZ{\mathbb{Z}}
\def\IR{\mathbb{R}}

\def\sd{{\bf d}}

\def\sx{{\bf x}}

\def\sJ{{\bf J}}
\def\se{{\bf e}}
\def\CK{{\cal K}}
\def\IF{{\cal F}}
\def\N{{\cal N}}
\def\bb{\bar{b}}

\def\bOmega{\bar{\Omega}}
\def\bpartial{\bar{\partial}}

\def\s*{\boldsymbol{*}}

\title{On the correspondence between D-branes and stationary supergravity solutions
of type II Calabi-Yau compactifications.\footnote{talk presented
at the Workshop on Strings, Duality and Geometry, CRM Montreal,
March 2000}}

\author{Frederik Denef\\ Department of Mathematics, Columbia University\\
New York, NY 10027, USA\\ \email{denef@math.columbia.edu}}

\abstract{In this talk, I review how four dimensional stationary
supergravity solutions that are more general than spherically
symmetric black holes emerge naturally in the low energy
description of BPS states in type II Calabi-Yau compactifications.
An explicit construction of multicenter solutions using single
center attractor flows as building blocks is presented, and some
interesting properties of these solutions are examined. We end
with a brief remark on non-BPS configurations. }

\begin{document}

\section{Introduction}

Calabi-Yau compactifications of type II string theory, which have
${\cal N}=2$ residual supersymmetry in four dimensions, are known
to have a moduli dependent spectrum of wrapped BPS D-branes; such
branes, observed as BPS particles in four dimensions, can for
example decay at so-called surfaces of marginal stability, similar
to the well known decays of BPS particles in Seiberg-Witten theory
\cite{SW}. In type IIB theory, when the geometric D-brane picture
can be trusted, the mathematical equivalent of existence of a BPS
brane in a certain homology class, is the existence of a special
lagrangian submanifold in that class. In recent years, the latter
problem got quite some attention, though it turned out to be an
extremely difficult issue \cite{joyce,hitchin}, and the general
existence question remains largely unsolved. In type IIA theory,
the mathematical equivalent is the existence of certain
holomorphic bundles on holomorphic submanifolds, again when the
geometric D-brane picture can be trusted. This problem is better
under control, though our understanding is far from complete.
Existence of BPS states in stringy regimes of the moduli space,
such as the Gepner point of the Quintic, can in favorable
circumstances be tackled from a pure conformal field theory
perspective, building on the work of \cite{RS}. Substantial work
in this context has been pioneered in \cite{BDLR} and
significantly extended in \cite{D,DR,Sch,BS,DFR,DFR2,DD,FM,KM}.

From a complementary, four dimensional low energy point of view,
one expects to have BPS solutions to the supergravity equations of
motion for any BPS state in the spectrum, at least if supergravity
can be trusted. The simplest solutions of this kind are
spherically symmetric black holes. Those were first studied in
${\cal N}=2$ theories in \cite{FKS}, where it was shown that they
exhibit a remarkable ``attractor'' feature: the value of the
moduli at the horizon are fixed by the charge of the black hole,
in the sense that they are invariant under continuous changes of
the moduli at infinity. In \cite{M}, where this property was
linked to a vast and still largely unexplored treasure of
arithmetical properties, it was noted that the existence of these
solutions is a nontrivial problem, depending strongly on the value
of the charges and vacuum moduli. It is therefore natural to
conjecture \cite{M} a correspondence between the existence of
those supergravity solutions and the existence of BPS D-brane
states in the full string theory.

However, as pointed out in \cite{D,branessugra}, this conjecture
fails in a number of established cases, where the state is known
to exist in string theory, but the corresponding black hole
solution does not exist in the supergravity theory, even in
regimes where supergravity can clearly be trusted. Obviously, from
the physics point of view, this is a major consistency problem.

The solution to this paradox was discovered in \cite{branessugra}:
the restriction to spherically symmetric black holes turned out to
be too narrow. For one, solutions corresponding to branes wrapped
around conifold cycles, though still spherically symmetric, are
{\em not} black holes, but rather ``empty holes'', as a result of
a mechanism reminiscent of the enhan\c{c}on mechanism
of~\cite{enhancon}, with a core carrying instead of a massless
vector multiplet and enhanced gauge symmetry, a massless
hypermultiplet. But more importantly, ${\cal N}=2$ supergravity
allows for solutions involving mutually nonlocal charges at rest
at a finite equilibrium distance from each other. Those solutions
are in general  stationary but non-static, as they can carry a
(quantized) intrinsic angular momentum, much like the
monopole-electron system in ordinary Maxwell theory. And as it
happens, some of the BPS states found in string theory can only be
realized in the low energy theory as such multi-center solutions.

General stationary solutions of four dimensional ${\cal N}=2$
supergravity were first explored in \cite{BLS}, further analyzed
from a geometrical point of view in \cite{branessugra}, and given
a rigorous, systematic treatment, including $R^2$ corrections, in
\cite{CWKM}.

The multicenter configurations also give a beautiful low energy
picture of what happens at marginal stability: the state can
literally be seen to decay smoothly into its constituents.
Furthermore, Joyce's stability criterion for special lagrangian
submanifolds \cite{joyce} is elegantly recovered and generalized,
and a strong similarity to the general Pi-stability criterion
proposed in \cite{DFR} emerges.

This opens up the exciting possibility that the existence
conjecture of \cite{M}, suitably generalized to include
multicenter solutions, should be taken seriously. The consequences
for mathematics and string theory of this correspondence, provided
it is true, are clearly far reaching. For example, it would enable
us to study the D-brane spectrum of compact Calabi-Yau
compactifications (and its mathematical equivalents), in a quite
systematic way, a problem that has been pretty elusive thus far
using other approaches. This issue, for type IIA theory on the
Quintic, will be addressed in a forthcoming paper \cite{quintic},
mainly from a numerical perspective.

In this talk, I will review how solutions more general than
spherically symmetric black holes arise in the low energy
description of BPS states, focusing on the solutions to the BPS
equations rather than on the technicalities of their derivation,
for which we refer to \cite{branessugra}. A detailed construction
of multicenter solutions from single center flows, a closer
examination of some properties of these solutions and a brief
digression on non-BPS states, extend the results of this
reference.

\section{Geometry of IIB/CY compactifications}
\label{sec2}

To establish our notation and setup, let us briefly review the low
energy geometry of type-IIB string theory compactified on a
Calabi-Yau 3-fold. We will always work in the type IIB framework,
but the equivalence with type IIA through mirror symmetry will be
implicitly assumed in the presentation of our examples.

We will follow the manifestly duality invariant formalism
of~\cite{M}. Consider type-IIB string theory compactified on a
Calabi-Yau manifold $X$. The four-dimensional low energy theory is
$\N = 2$ supergravity coupled to $n_v = h^{1,2}$ massless abelian
vectormultiplets and $n_h = h^{1,1}+ 1$ massless hypermultiplets,
where the $h^{i,j}$ are the Hodge numbers of $X$. The
hypermultiplet fields will play no role in the following and are
set to zero.

The vectormultiplet scalars are given by the complex structure
moduli of $X$, and the lattice of electric and magnetic charges is
identified with $H^3(X,\IZ)$, the lattice of integral harmonic
$3$-forms on $X$. The ``total'' electromagnetic field strength
$\IF$ is (up to normalisation convention) equal to the type-IIB
self-dual five-form field strength, and is assumed to have values
in $\Omega^2(M_4) \otimes H^3(X,\IZ)$, where $\Omega^2(M_4)$
denotes the space of 2-forms on the four-dimensional spacetime
$M_4$. The usual components of the field strength are retrieved by
picking a symplectic basis ${\alpha^I,\beta_I}$ of $H^3(X,\IZ)$:
\begin{equation}
\IF = F^I \otimes \beta_I - G_I \otimes \alpha^I\,. \label{FGcomp}
\end{equation}
A 3-brane wrapped around a cycle Poincar\`e dual to $\Gamma \in
H^3(X,\IZ)$ has electric and magnetic charges equal to its
components with respect to this basis. The total field strength
satisfies the self-duality constraint: $ \IF = *_{10} \IF\ $,
which translates to electric-magnetic duality in the four
dimensional theory.

The geometry of the vector multiplet moduli space, parametrized
with $n_v$ coordinates $z^a$, is special K\"ahler~\cite{SG}. The
(positive definite) metric
\begin{equation} \label{SKmetric}
g_{a\bb} = \partial_a \bpartial_{\bb} \CK
\end{equation}
is derived from the K\"ahler potential
\begin{equation}
\CK = - \ln \left( i \int_X \Omega_0 \wedge \bOmega_0 \right),
\end{equation}
where $\Omega_0$ is the holomorphic $3$-form on $X$, depending
holomorphically on the complex structure moduli. It is convenient
to introduce also the {\em normalized} 3-form
\begin{equation} \label{Omdef}
\Omega = e^{\CK/2} \, \Omega_0\,.
\end{equation}
The ``central charge'' of $\Gamma \in H^3(X,\IZ)$ is given by
\begin{equation} \label{Zdef}
Z(\Gamma) \equiv \int_X \Gamma \wedge \Omega \equiv \int_\Gamma
\Omega\,,
\end{equation}
where we denoted, by slight abuse of notation, the cycle
Poincar\'e dual to $\Gamma$ by the same symbol $\Gamma$. Note that
$Z(\Gamma)$ has nonholomorphic dependence on the moduli through
the K\"{a}hler potential.

We will make use of the (antisymmetric, topological, moduli
independent) intersection product:
\begin{equation} \label{intproddef}
\langle \Gamma_1,\Gamma_2 \rangle = \int_X \Gamma_1 \wedge
\Gamma_2 = \#(\Gamma_1 \cap \Gamma_2)\,.
\end{equation}
With this notation, we have for a symplectic basis $\{
\alpha^I,\beta_I \}$ by definition $\langle \alpha^I,\beta_J
\rangle = \delta^I_J$. Integrality of this intersection product is
equivalent to Dirac quantization of electric and magnetic charges.

\section{BPS equations of motion}
\label{BPSeom}

\subsection{The static, single center, spherically symmetric case}
\label{BPSeomstatic}

The BPS equations of motion for the static, spherically symmetric
case were derived in \cite{FKS}, and cast in the form of first
order flow equations on moduli space in \cite{FGK}. We assume a
charge $\Gamma \in H^3(X,\IZ)$ is located at the origin of space.
The spacetime metric is of the form
\begin{equation} \label{staticmetric}
 ds^2 = - e^{2U} dt^2 + e^{-2 U} dx^i
dx^i \,,
\end{equation}
with $U$ a function of the radial coordinate distance
$r=|\bf{x}|$, or equivalently of the inverse radial coordinate
$\tau=1/r$. The BPS equations of motion for $U(\tau)$ and the
moduli $z^a(\tau)$ are:
\begin{eqnarray}
 \partial_\tau U &=& - e^U |Z| \,,\label{at1} \\
 \partial_\tau z^a &=& -2 e^U g^{a\bb} \, \bpartial_{\bb} |Z|\,, \label{at2}
\end{eqnarray}
where $Z=Z(\Gamma)$ is as in (\ref{Zdef}) and $g_{a\bb}$ as in
(\ref{SKmetric}). A closed expression for the electromagnetic
field, given the solutions of these flow equations, can be found
e.g.\ in \cite{branessugra}.

This is the form of the BPS equations found in \cite{FGK}. An
alternative form of the equations, essentially equivalent to those
found in \cite{FKS}, is:
\begin{equation} \label{bps3}
2 \, \partial_\tau \left[ e^{-U} \im \left(e^{-i \alpha} \Omega
\right) \right] = - \Gamma\,,
\end{equation}
where $\alpha = \arg Z$, which can be shown to be the phase of the
conserved supersymmetry (see e.g. \cite{CWKM}). Note that this
nice compact equation actually has $2 n_v + 2$ real components,
corresponding to taking intersection products with the $2 n_v + 2$
elements of a basis $\{C_L\}_L$ of $H^3(X,\IZ)$:
\begin{equation} \label{bpscomp}
2 \, \partial_\tau \left[ e^{-U} \im \left( e^{-i \alpha} Z(C_L)
\right) \right] = - \langle C_L , \Gamma \rangle \,,
\end{equation}
One component is redundant, since taking the intersection product
of (\ref{bps3}) with $\Gamma$ itself produces trivially $0=0$.
This leaves $2 n_v + 1$ independent equations, matching the number
of real variables $\{U,\re z^a,\im z^a\}$.

Note that alternatively, we could have left $\alpha$ as an
arbitrary field instead of putting it equal to $\arg Z$. Then the
previously redundant component of the equation gives $\im(e^{-i
\alpha} Z)=0$, hence $\alpha=\arg Z$ or $\alpha=\arg(-Z)$. The
latter possibility is automatically excluded however, since it
gives rise to a highly singular, unphysical solution. Indeed, this
case corresponds to (\ref{at1})-(\ref{at2}) with the sign of the
right hand sides reversed. Then $|Z|$ and $e^U$ would be
increasing functions in $\tau$, with $e^U$ satisfying the estimate
$e^U \geq \frac{e^{U(\tau_0)}}{1-e^{U(\tau_0)} |Z(\tau_0)|
(\tau-\tau_0)}$ for any $\tau_0$ and $\tau>\tau_0$. Since this
diverges at finite $\tau$, the solution breaks down. Note that
this candidate solution in any case would have had negative ADM
mass and be gravitationally repulsive --- physically quite
undesirable properties. So only the possibility $\alpha=\arg Z$
remains, bringing us back to the original setup of the equations.

Since the right hand side of (\ref{bpscomp}) consists of
$\tau$-independent integer charges, (\ref{bps3}) readily
integrates to
\begin{equation} \label{integrated}
 2\, e^{-U} \im \left(e^{-i \alpha} \Omega\right)= -\Gamma \, \tau +
 2  \im \left(e^{-U} e^{-i \alpha} \Omega\right)_{\tau=0}.
\end{equation}
For asymptotically flat space, $U_{\tau=0}=U_{r=\infty}=0$.

In contrast to \cite{FKS} and most of the older attractor
literature, we prefer to work with the {\em normalized} periods
and an explicit phase factor $e^{i \alpha}$. The difference
amounts to nothing more than a normalization gauge choice, but it
proves to be conceptually more transparent, and
numeric-computationally far more convenient to make the above
choice.

The result (\ref{integrated}) is very powerful, as it {\em solves}
in principle the equations of motion. Of course, finding the
explicit flows in moduli space from~(\ref{integrated}) requires
inversion of the periods to the moduli, which in general is not
feasible analytically. However, in large complex structure
approximations or numerically for e.g.\ the quintic, this turns
out to be possible.

There is one catch to (\ref{integrated}) though, namely, as shown
in \cite{branessugra}, it is not valid for a vanishing cycle
$\Gamma$, at values of the moduli where it vanishes. Indeed,
looking for example at a one modulus case \footnote{where we take
the modulus $z$ to be the unnormalized holomorphic $\Omega_0$
period, say.} near a conifold point, we see that while
(\ref{at1})-(\ref{at2}) allows for solutions with constant $z$ and
$U$ at the conifold point (since the inverse metric becomes zero
there), this is not the case for (\ref{integrated}). The correct
equation in this case is (\ref{at1})-(\ref{at2}). This subtlety is
important in the discussion of ``empty hole'' solutions (see
below), and it thus eliminates some confusion in the older
attractor literature, where solutions with charge corresponding to
a conifold cycle seemed to emerge that were very unphysical (naked
curvature singularities, gravitational repulsion, etc.). Such
pathological behavior is indeed what one gets when naively
applying (\ref{integrated}) to those cases.

Finally, it was also observed in \cite{branessugra} that the BPS
equations of motion can be interpreted as geodesic equations for
stretched strings with varying tension in a certain curved
background, making contact, at least in a rigid (gravity
decoupling) limit of the theory, with the ``3-1-7'' brane picture
of BPS states in ${\cal N}=2$ quantum field theory
\cite{BPS37,threeprong}.

\subsection{The general stationary case}

The BPS equations of motion for the general case, though of course
more complicated, are quite similar in structure to those of the
single center case. For a derivation, we refer to
~\cite{branessugra} and ~\cite{CWKM}. In the latter reference, the
equations below were shown to describe the most general stationary
BPS solutions, provided a certain ansatz was made for the
embedding of the residual supersymmetry. More general solutions
could exist, but a fully general analysis proved to be too
cumbersome to be carried out thus far.

The metric will be of the form
\begin{equation} \label{stationarymetric}
 ds^2 = - e^{2U} \left(dt + \omega_i dx^i\right)^2 + e^{-2 U} dx^i
 dx^i\,,
\end{equation}
where $U$ and $\omega$, together with the moduli fields $z^a$, are
time-independent solutions of the following BPS equations,
elegantly generalizing (\ref{integrated}):
\begin{eqnarray}
 2 \, e^{-U} \im \left(e^{-i\alpha} \Omega\right) &=& H \,, \label{mc1}\\
 \s* d \omega &=& \langle d H,H \rangle \,, \label{mc2}
\end{eqnarray}
with $H(\sx)$ an $H^3(X)$-valued harmonic function (on flat
coordinate space $\IR^3$), and $\s*$ the flat Hodge star operator
on $\IR^3$. For $N$ charges $\Gamma_p$ located at coordinates
$\sx_p$, $p=1,\ldots,N$, in asymptotically flat space, one has:
\begin{equation}
  H = -\sum_{p=1}^N \Gamma_p \, \tau_p  \, \, + \,  2 \im\left(e^{-i
  \alpha} \Omega\right)_{r=\infty} \,, \label{Haha}
\end{equation}
with $\tau_p=1/|\sx-\sx_p|$.

Note that at large $r$, equation (\ref{mc1}) reduces to lowest
order in $1/r$ to the spherically symmetric case
(\ref{integrated}). The phase $\alpha$ in (\ref{mc1}) is to be
considered as an a priori independent unknown field here. However,
a reasoning similar to the comments under (\ref{bpscomp}) shows
that it must be asymptotically equal to the phase of $Z(\sum_p
\Gamma_p)$ for $|\sx| \to \infty$.\footnote{This will be made more
explicit in section \ref{multicentersol}. \label{fnb}}

The electromagnetic field is again determined algebraically from
the solutions of the BPS equations for $U$, $\omega$ and $z^a$. We
refer to \cite{branessugra} (or \cite{CWKM}) for the explicit
expressions.

\section{Solutions}
\label{solutions}

\subsection{Static single center case:
black, empty and no holes.} \label{ben}

In asymptotically flat space (which we will assume unless stated
otherwise), all well-behaved solutions to
to~(\ref{at1})--(\ref{at2}) saturate the BPS bound $M_{ADM} =
|Z_{\tau=0}|$. Another simple universal characteristic is that the
``size'' of the solutions scales proportional to the charge
number, due to a trivial rescaling symmetry of the equations of
motion. An important and less trivial general property is the
following. Equation (\ref{at2}) implies $\partial_\tau |Z| = -4
e^{U} g^{a\bb} \,
\partial_a |Z| \, \bpartial_{\bb} |Z| \leq 0$. Therefore the flows
in moduli space for increasing $\tau$, given by the BPS equations,
will converge to minima of $|Z|$, and the corresponding moduli
values are generically invariant under continuous deformations of
the moduli at spatial infinity, so they only depend on the charge
$\Gamma$, a phenomenon referred to as the attractor mechanism. One
distinguishes three cases \cite{M}, depending on the value of the
minimum of $|Z|$ (zero or nonzero) and its position in moduli
space (at singular or regular point):
\begin{enumerate}
 \item {\em nonzero minimal $|Z|$}. This yields a regular BPS black
 hole. The near-horizon geometry is $AdS_2 \times S^2$, and the
 horizon area equals $4 \pi |Z|^2_{min}$. From the BPS equations
 of motion, one directly deduces that at the horizon,
 $2 \im(\bar{Z} \Omega) = - \Gamma$. This equation, determining
 (locally) the position of the attractor point in moduli space, is
 often called the attractor equation.

 \item {\em zero minimal $|Z|$ at a regular point in moduli space}. In
 this case, the BPS equations do not have a solution \cite{M}:
 $Z=0$ will be reached by the flow at a finite radius, beyond which the
 solution cannot be continued. This is consistent with physical
 expectations: in a vacuum {\em close} to a regular zero of $Z$ in moduli
 space, the state cannot exist, as it would imply the existence of
 a massless particle at the zero locus, which in turn should create a
 singularity in moduli space \cite{S}, in contradiction with the
 supposed regularity of the point under consideration.

 \FIGURE[t]{\centerline{\epsfig{file=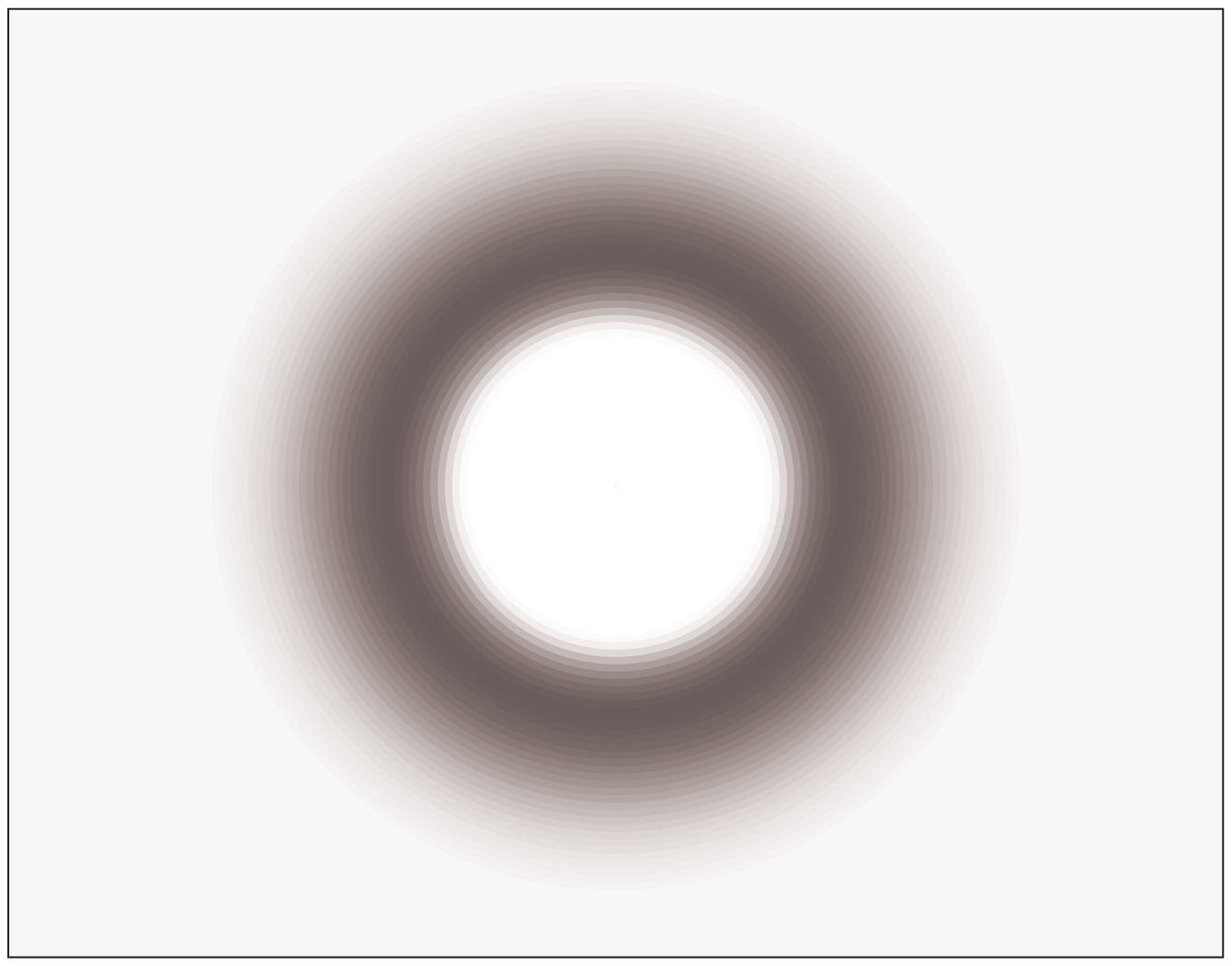,height=5cm}}
 \caption{Energy density sketch of an empty hole.}\label{empty}}

 \item {\em zero minimal $|Z|$ at a singular or boundary
 point in moduli space}. In
 this case, the BPS equations may or may not have a solution. In
 the case of a conifold cycle for example, the equations do have a
 solution, describing an ``empty hole'' \cite{branessugra}: again,
 the zero of $Z$ (i.e. the conifold locus in moduli space)
 is reached at finite radius, but now, as mentioned at the end of
 section \ref{BPSeomstatic}, the solution {\em can} be
 continued in a ($1 \times$) continuous differentiable way,
 simply as flat space (i.e.\ constant $U$) with the
 moduli fixed at the conifold locus. This is illustrated in fig.\
 \ref{empty}. Inside this core, a test
 conifold particle would be massless, and the charge source
 becomes completely delocalized. The latter is illustrated for
 example by the fact that the core radius of an emty hole
 increases when the background moduli approach the conifold radius
 (since the flow reaches the attractor point ``earlier'' in
 $\tau$). Therefore, if we let another conifold particle approach our
 initial empty hole, the radius of that particle will increase,
 eventually smoothly ``melting'' into the original core (this process
 can in fact be analyzed quantitively by considering multicenter
 solutions with parallel charges).

 All this is quite similar to the
 enhan\c{c}on mechanism of \cite{enhancon}, the main difference
 being that it is a hypermultiplet becoming massless in the core
 here instead of a vector multiplet. The analogon of the
 unphysical, naive ``repulson'' solution of \cite{enhancon} is the
 naive (and wrong) solution one would get by continuing
 (\ref{integrated}) inside the core.

 Of course, from the full
 string theoretic point of view, we cannot necessarily
 trust the usual low energy supergravity lagrangian all the way close to
 the conifold
 locus, because in principle there is an additional (nearly) massless field to
 be included. However, for the four dimensional supergravity
 theory on itself, the empty hole solutions are perfectly well behaved, and
 exhibit some properties that are physically very pleasing for such
 states \cite{branessugra}, such as the absence of a horizon,
 and slow motion scattering (probably) without the coalescence effect typical
 for black holes \cite{modulispace}. Indeed, one expects 3-branes wrapped
 around a conifold cycle to behave like elementary particles that
 can be consistently decoupled from gravity, rather than as black
 holes, and one does not expect bound states of several copies of
 such branes \cite{S}. It is quite nice to see this emerging from the low
 energy description here.

 It would perhaps be interesting to find out whether empty holes, like
 their black hole cousins~\cite{adsfrag}, also have some sort of
 a Maldacena dual~\cite{AdSCFT} QFT description.

\end{enumerate}

\subsection{Configurations with uniformly charged spherical shells}

\subsubsection{Static equilibrium and marginal stability surfaces}
\label{stateq}

There is a relatively simple but nontrivial generalization of
these spherically symmetric solutions, which in the end will
provide one way out of the paradoxes mentioned in the
introduction, namely configurations involving one or more
uniformely charged spherical shells (see fig. \ref{shells}). In
\cite{branessugra} it was explained how such configurations can
arise naturally in a physical process.

\FIGURE[t]{\epsfig{file=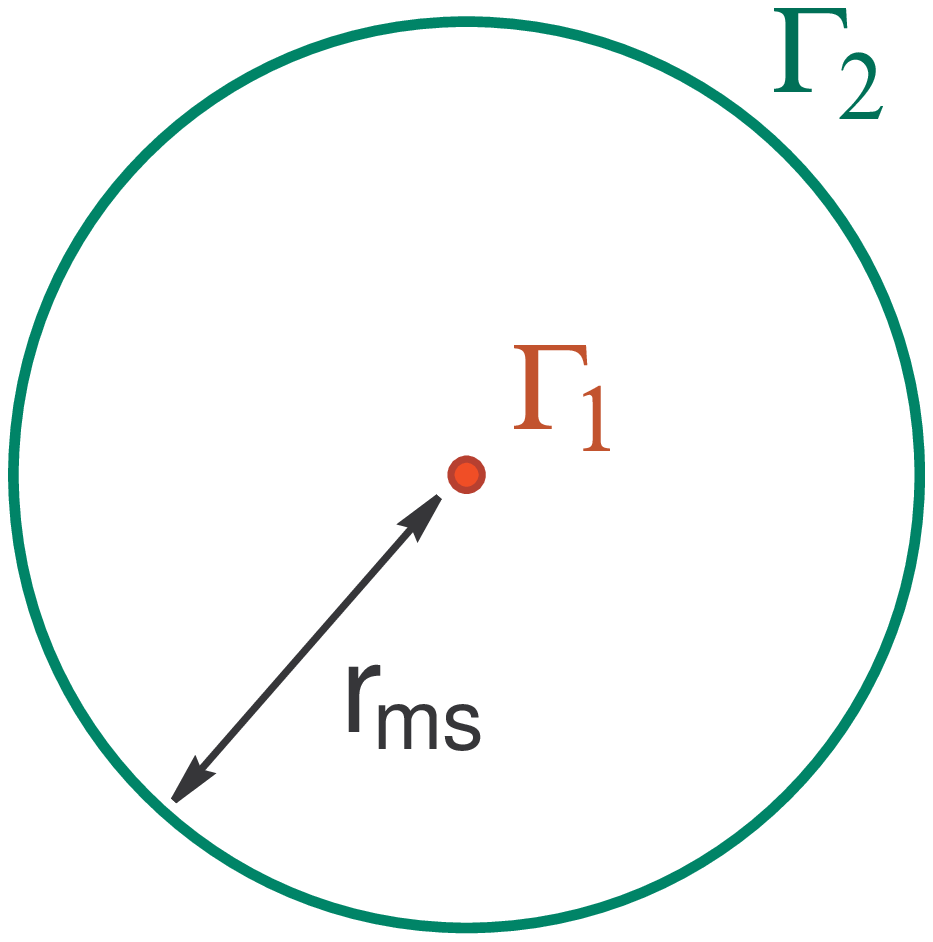,height=5cm}
 \caption{A configuration consisting of a uniformly
 charged spherical shell of charge $\Gamma_2$ surrounding a charge $\Gamma_1$
 centered at the origin. $r_{\mathrm{ms}}$ is the coordinate BPS equilibrium distance.}
 \label{shells}}

The BPS solution of the field equations between two shells are
identical to the usual spherical symmetric ones, with the
appropriate enclosed charge substituted, and the solutions in
adjacent regions matched by continuity. However, to get a complete
(and stable) BPS configuration, the various energy contributions
(energy stored in fields and ``bare'' mass of the shells) should
precisely add up to $|Z(\Gamma)|_{r=\infty}$, with $\Gamma$ the
total charge. Let us for example consider the one shell case of
fig. \ref{shells}. Denote the radius of the shell by
$r_{\mathrm{ms}}$. The energy in the bulk fields outside the
$\Gamma_2$-shell can be seen to be
$E_{\mathrm{out}}=|Z(\Gamma)|_{r=\infty}-(e^U
|Z(\Gamma)|)_{r=r_{\mathrm{ms}}}$, with
$\Gamma=\Gamma_1+\Gamma_2$. The bare energy of the shell itself is
$E_{\mathrm{shell}}=(e^U |Z(\Gamma_2)|)_{r_{\mathrm{ms}}}$. The
energy inside the shell is $E_{\mathrm{in}}=(e^U
|Z(\Gamma_2)|)_{r_{\mathrm{ms}}}$. So the total energy is
\begin{equation} \label{bpscond}
E_{\mathrm{tot}}=|Z(\Gamma)|_{\infty}+ \left( e^U
(|Z(\Gamma_1)|+|Z(\Gamma_2)|-|Z(\Gamma_1)+Z(\Gamma_2)|)
\right)_{r_{\mathrm{ms}}}.
\end{equation}
To saturate the BPS bound, the second term must vanish. This is
the case if and only if the phases of $Z(\Gamma_1)$ and
$Z(\Gamma_2)$ are equal for the values of the moduli at
$r=r_\mathrm{ms}$, that is, if the flow in moduli space given by
the solution crosses a surface of $(\Gamma_1,\Gamma_2)$ marginal
stability at $r=r_{\mathrm{ms}}$ (explaining the subscript ``ms''
for this radius).

Because of the BPS condition, these configurations can be expected
to be stable. To verify this, one can compute the force potential
$W$ on a test particle of charge $\Gamma_t$ at rest in the
background (BPS) field of a charge $\Gamma_0$, starting from the
DBI+WZ action for a D-brane in an external field
\cite{branessugra}. The result is:
\begin{equation} \label{potexpr}
W(r)= \left. 2 \, e^U \, |Z(\Gamma_t)| \, \sin^2 (
\frac{\alpha_t-\alpha_0}{2} ) \, \, \right|_r \,,
\end{equation}
where $\alpha_i = \arg Z(\Gamma_i)$. This potential is everywhere
positive, and acquires a zero minimum when
$\alpha_t(r)=\alpha_0(r)$, that is, indeed, at marginal stability.
A specific example is shown in fig. \ref{potential}.

\FIGURE[t]{\centerline{\epsfig{file=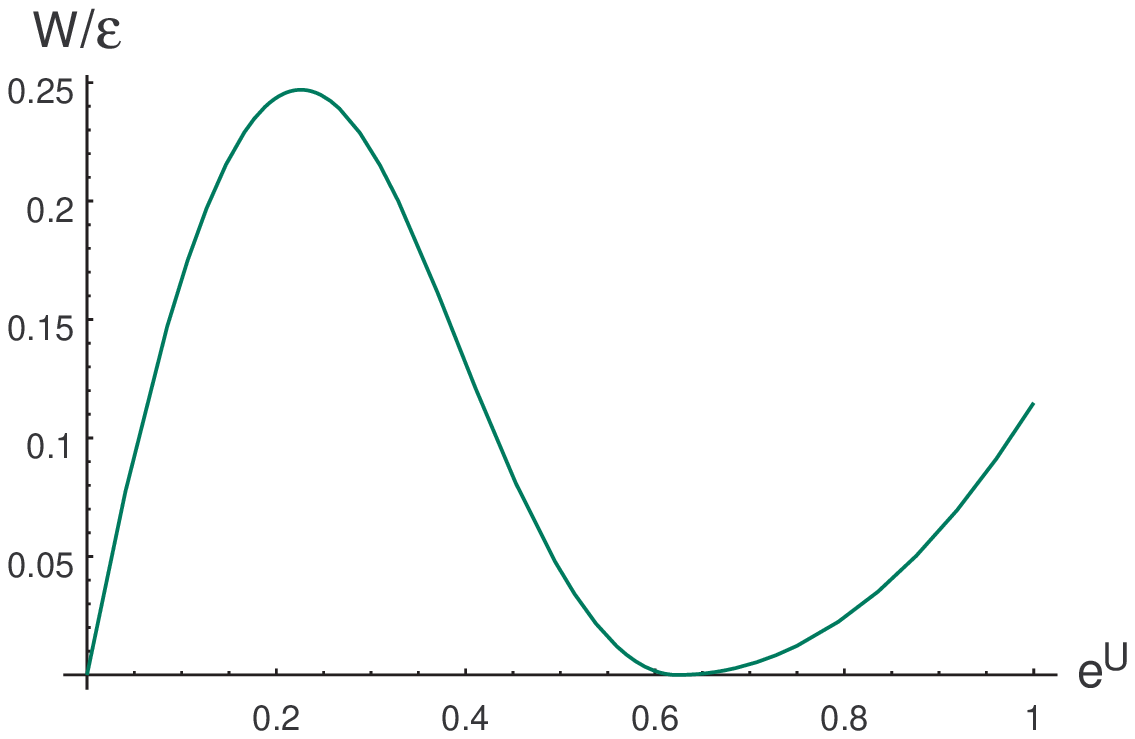,height=5cm}}
\caption{Example of a force potential for a test particle with
charge $\epsilon \, \Gamma_2$ at rest in the background field of a
BPS configuration like the one shown in fig. \ref{shells}. As a
convenient radial coordinate, we use the redshift factor $e^{U}$
(which increases monotonically from 0 to 1 when going from horizon
to spatial infinity). For this particular example, we computed the
potential numerically, using mirror symmetry and \cite{GLaz}, for
a type IIA compactification on the quintic, with, in the
conventions of \cite{BDLR},
$(Q_{D6},Q_{D4},Q_{D2},Q_{D0})(\Gamma_1) = (1,2,5,-7)$,
$(Q_{D6},Q_{D4},Q_{D2},Q_{D0})(\Gamma_2) = (-1,1,4,-1)$, and
vacuum modulus $\psi=7 e^{i \pi/5}$.}\label{potential}}

\subsubsection{Closed expression for equilibrium distance, and existence of solutions}

A closed expression for the equilibrium radius $r_{\mathrm{ms}}$
can be extracted from the integrated flow
equation~(\ref{integrated}) for the fields outside the shell.
Taking the intersection product of $\Gamma_1$ with this equation
gives, denoting $Z(\Gamma_i)$ in short as $Z_i$:
\begin{equation} \label{imZ1}
2 \im (e^{-U} e^{-i \alpha} Z_1) = - \langle \Gamma_1,\Gamma
\rangle \, \tau  +  2 \, \im (e^{-i \alpha} Z_1)_{\tau=0} \,.
\end{equation}
At $1/\tau=r=r_{\mathrm{ms}}$, the left-hand side is zero, so
\begin{equation} \label{prermsformula}
r_{\mathrm{ms}} = \frac{\langle \Gamma_1,\Gamma \rangle}{2 \,
\im(e^{-i \alpha} Z_1)_{r=\infty}}\,.
\end{equation}
Using $e^{i \alpha}=Z/|Z|$ with $Z=Z_1+Z_2$ and $\langle
\Gamma_1,\Gamma \rangle = \langle \Gamma_1,\Gamma_2 \rangle$, this
can be written more symmetrically as
\begin{equation} \label{rmsformula}
r_{\mathrm{ms}} = \frac{1}{2} \langle \Gamma_1,\Gamma_2 \rangle
\left. \frac{|Z_1+Z_2|}{\im(\bar{Z_2} Z_1)} \right|_{r=\infty}.
\end{equation}

Such composite configurations will not exist for all possible
charge combinations ($\Gamma_1,\Gamma_2$) in a given vacuum. For
example, a necessary condition for existence is obviously
$r_{\mathrm{ms}}>0$, with $r_{\mathrm{ms}}$ given by charge and
vacuum data as in (\ref{rmsformula}).\footnote{In particular,
$\Gamma_1$ and $\Gamma_2$ should be mutually nonlocal.
\label{fna}} This is not a sufficient condition however. For
example, the flow could hit a zero before it reaches the surface
of $(\Gamma_1,\Gamma_2)$ marginal stability. Or the unique point
in the flow where the left-hand side of (\ref{imZ1}) is zero,
could correspond to a point in moduli space where $Z_1$ and $Z_2$
have {\em opposite} instead of equal phases. Furthermore, of
course, $\Gamma_2$ has to exist as a BPS state for the values of
the moduli at $r_{ms}$, and so should the solution inside the
shell.

\subsection{General, multicenter, stationary case} \label{multicentersol}

The above spherical solutions, while interesting and suggestive,
are still not at the same level as genuine black hole soliton
solutions of supergravity, in the sense that we explicitly added a
smeared out charge source with a nonvanishing bare mass
contribution to the total energy. On the other hand, the
expression (\ref{potexpr}) for the potential of a test particle
suggests the existence of truly solitonic BPS solutions with only
point charges, located at equilibrium distance ($r_{\mathrm{ms}}$
in fig. \ref{shells}) from each other. Such solutions, in the
limit of a large number of charges positively proportional to
$\Gamma_2$, evenly distributed over a sphere at
$r=r_{\mathrm{ms}}$ from a black hole center with charge
$\Gamma_1$, can be expected to approach the spherically symmetric
case away from $r=r_{\mathrm{ms}}$. Sufficiently close to the
$\Gamma_2$ charges on the other hand, the solution can be expected
to approach a pure $\Gamma_2$ black hole solution.

To address this problem quantitavely, we should look for solutions
to the general multicenter BPS equations given by
(\ref{mc1})--(\ref{Haha}). Because of the remark in footnote
\ref{fna}, we expect the relevant solution to involve mutually
nonlocal charges. The latter complicates the situation
considerably, since such configurations will in general not be
static, because the right hand side of (\ref{mc2}) is nonvanishing
and hence $\omega$ cannot be gauged away.

\subsubsection{Properties of muticenter solutions with mutually nonlocal charges}

Assuming we have a solution to those equations, let us see what
properties we can deduce. A first observation is that there will
be constraints on the positions of the charges. Indeed, acting
with $\sd \s*$ on equation~(\ref{mc2}) gives
\begin{equation}
0 = \langle \Delta H, H \rangle \,,
\end{equation}
with $\Delta$ the (flat) laplacian on $\IR^3$, so,
using~(\ref{Haha}) and $\Delta \tau_p = -4 \pi
\delta^3(\sx-\sx_p)$, we find that for all $p=1,\dots,N$:
\begin{equation} \label{distconstr}
\sum_{q=1}^N \frac{\langle \Gamma_p,\Gamma_q
\rangle}{|\sx_p-\sx_q|} = 2 \, \im\left(e^{-i \alpha}
Z(\Gamma_p)\right)_{r=\infty}.
\end{equation}
Note that the full moduli space of solutions to the constraints
(\ref{distconstr}) will have a fairly complicated structure.
However, in the particular case of one source with charge
$\Gamma_1$ at $\sx=0$ and $M$ sources with charges postively
proportional to $\Gamma_2$ at positions $\sx_p$, the constraints
simplify to
\begin{equation}
|\sx_p| = \frac{\langle \Gamma_2,\Gamma_1 \rangle}{2 \, \im(e^{-i
\alpha} Z(\Gamma_2))_{r=\infty}}\,,
\end{equation}
which is, as expected, precisely the equilibrium distance
$r_{\mathrm{ms}}$ found in the spherical shell picture,
equation~(\ref{rmsformula}).

Incidentally, by summing equation (\ref{distconstr}) over all $p$,
one gets $\im(e^{-i\alpha} Z(\Gamma))_{\infty}=0$, with
$\Gamma=\sum_p \Gamma_p$. On the other hand, by taking the
intersection product of (\ref{mc1}) with any $\Gamma_p$, and using
(\ref{distconstr}), one sees that in the limit $\sx \to \sx_p$,
that is, $\tau_p \to \infty$ and $\tau_q \to 1/|\sx_p-\sx_q|$ for
$q \neq 0$, one has $\im(e^{-i \alpha} Z(\Gamma_p)) \to 0$.
Therefore we have
\begin{equation} \label{alphacondition}
 \alpha_{|\sx|=\infty} = \arg Z(\Gamma)_{\infty} \qquad \mbox{ and }
 \qquad \alpha_{\sx = \sx_p} = \arg Z(\Gamma_p)_{\sx_p} \, .
\end{equation}
As argued under (\ref{bpscomp}), the opposite sign for
$e^{i\alpha}$ is to be excluded, as it gives an unphysical and
severely singular negative mass solution, corresponding to a flow
in the ``wrong direction'' in moduli space \cite{branessugra}.

Note that this also implies that very far from all charges, as
well as very close (in terms of coordinate distance) to any one of
them, the solution will approach the single center case. Thus, if
the solution exists, we can expect its image in moduli space to
look like a fattened, ``split'' flow, as sketched in fig.\
\ref{fattened}. We will come back to this point in much more
detail in section \ref{solconstruct}.

\FIGURE[t]{\centerline{\epsfig{file=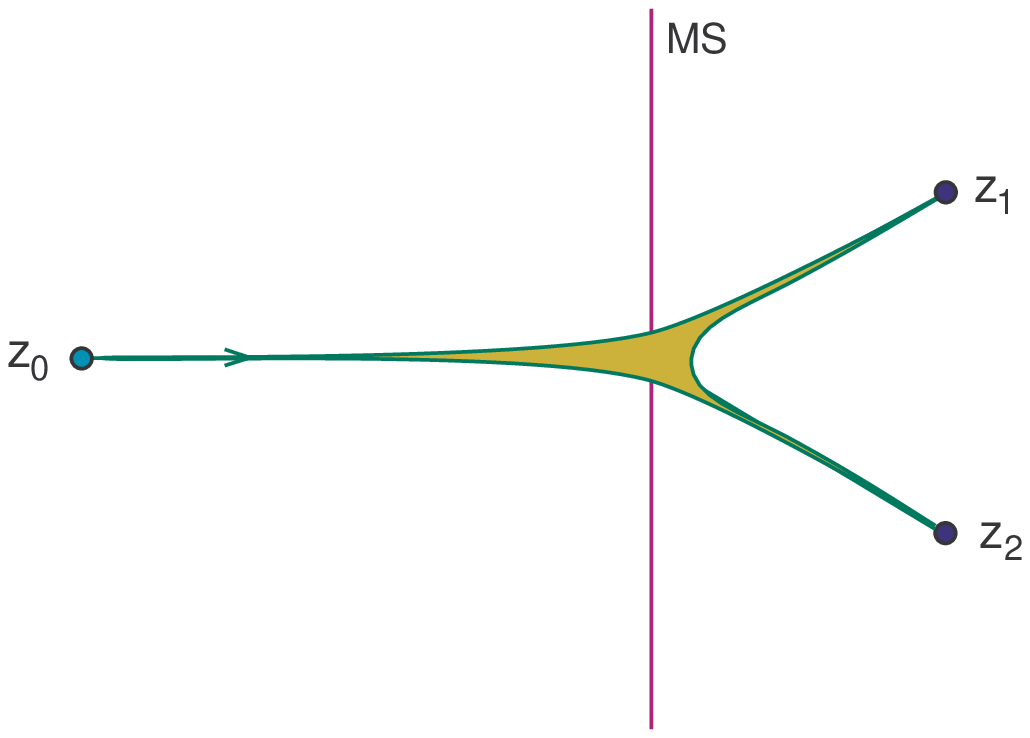,height=5cm}}
\caption{Sketch of the image of $z(\sx)$ in moduli space for a
multicenter solution containing two different charges $\Gamma_1$
and $\Gamma_2$, with attractor points $z_1$ resp.\ $z_2$, and
modulus at spatial infinity $z_0$. The line labeled ``MS'' is a
$(\Gamma_1,\Gamma_2)$-marginal stability line.}\label{fattened}}

A second property that can be deduced directly from the equations
is the total angular momentum of the solution. It is well known
from ordinary Maxwell electrodynamics that multicenter
configurations with mutually non-local charges (e.g.\ the
monopole-electron system) can have intrinsic angular momentum even
when the particles are at rest. The same turns out to be true
here.

We define the angular momentum vector $\sJ$ from the asymptotic
form of the metric (more precisely of $\omega$) as~\cite{MTW}
\begin{equation}
\omega_i = 2 \, \epsilon_{ijk} \, J^j \, \frac{x^k}{r^3}  +
O\left(\frac{1}{r^3}\right) \quad \mbox{for } r \longrightarrow
\infty \,.
\end{equation}
Plugging this expression in~(\ref{mc2}) and using~(\ref{Haha})
and~(\ref{distconstr}), we find, after some work,
\begin{equation}
\sJ = \frac{1}{2} \sum_{p<q} \langle \Gamma_p, \Gamma_q \rangle \,
{\se}_{pq} \,,
\end{equation}
where ${\se}_{pq}$ is the unit vector pointing from $\sx_q$ to
$\sx_p$:
\begin{equation}
{\se}_{pq} = \frac{\sx_p - \sx_q}{|\sx_p - \sx_q|} \,.
\end{equation}
Just like in ordinary electrodynamics, this is a ``topological''
quantity: it is invariant under continuous deformations of the
solution, and quantized in half-integer units (more precisely,
when all charges are on the z-axis, $2 J_z \in \IZ$). The
appearance of intrinsic configurational angular momentum implies
that quantization of these composites will have some non-trivial
features. In particular, when many particles are involved, the
ground state will presumably be highly degenerate.

\subsubsection{Construction and existence of solutions}
\label{solconstruct}

For simplicity, we will focus here on configurations with only two
different kinds of charges $\Gamma_1, \Gamma_2$, each distributed
over $N$ centers $x_{1,a}$, resp. $x_{2,a}$, $a=1,\ldots,N$. We
take $\Gamma_1$ and $\Gamma_2$ to be mutually nonlocal, $\langle
\Gamma_1,\Gamma_2 \rangle \neq 0$. Centers of equal charge may
coincide.

Then we can write the harmonic function $H$ of (\ref{Haha}) as
\begin{equation}
  H = - \Gamma_1 \, V_1(\sx) - \Gamma_2 \, V_2(\sx)  \, \, + \,  2 \im\left(e^{-i
  \alpha} \Omega\right)_{r=\infty} \,, \label{Haha2}
\end{equation}
with
\begin{equation}
  V_i(\sx) = \sum_{a=1}^{N} \frac{1}{|\sx - \sx_{i,a}|} \, ,
\end{equation}
and (\ref{distconstr}) becomes:
\begin{equation} \label{distconstr2}
V_{12} \equiv V_1(\sx_{2,b}) = V_2(\sx_{1,a}) = \frac{2}{\langle
\Gamma_1,\Gamma_2 \rangle} \left( \frac{\im (Z_1
\bar{Z_2})}{|Z_1+Z_2|} \right)_{r=\infty}.
\end{equation}

Taking the intersection product of (\ref{mc1}) with the source
charges $\Gamma_1$, $\Gamma_2$ and with a basis
$\{\Gamma^\perp_L\}_L$ of the vector space spanned by the elements
of $H^3(X,\IZ)$ which are local w.r.t.\ $\Gamma_1,\Gamma_2$ (i.e.,
they have zero intersection with both $\Gamma_1$ and $\Gamma_2$),
and using (\ref{distconstr2}), we get the equations
\begin{eqnarray} \label{bps2c1}
 2 \, e^{-U} \im[e^{-i \alpha} Z_1] &=& - \langle
 \Gamma_1,\Gamma_2 \rangle \, (V_2(\sx)-V_{12}) \,, \\
 \label{bps2c2}
2 \, e^{-U} \im[e^{-i \alpha} Z_2] &=& - \langle
 \Gamma_2,\Gamma_1 \rangle \, (V_1(\sx)-V_{12}) \,, \\
 \label{bps2c3}
2 \, e^{-U} \im[e^{-i \alpha} Z^\perp_L] &=& 2 \im[e^{-i
 \alpha} Z^\perp_L]_{r=\infty} \, ( = \mbox{const.}) \,.
\end{eqnarray}
This is a set of $2n+2$ independent equations, equivalent to
(\ref{mc1}), for $2n+2$ variables, where $n$ is the number of
moduli.

Similarly, the second BPS equation (\ref{mc2}) becomes
\begin{eqnarray}
 \s* \sd \, \omega &=& \langle \Gamma_1,\Gamma_2 \rangle
 [(V_2-V_{12}) \sd V_1 - (V_1-V_{12}) \sd V_2 ] \\
 &=& \langle \Gamma_1,\Gamma_2 \rangle \,
 (V_1-V_{12})(V_2-V_{12}) \, \sd \,
 \ln \left( \frac{V_1-V_{12}}{V_2-V_{12}} \right)
 \,.  \label{bps2c4}
\end{eqnarray}

We define two (local) space coordinate functions, $t$ and
$\theta$, as follows:
\begin{equation}
V_1(\sx) - V_{12} = t \cos \theta \quad ; \quad V_2(\sx) - V_{12}=
t \sin \theta \, ,
\end{equation}
with $t>0$. So
\begin{eqnarray}
  t &=& \sqrt{(V_1-V_{12})^2+(V_2-V_{12})^2} \, , \label{tdef}\\
  \tan \theta &=& \frac{V_1-V_{12}}{V_2-V_{12}} \, .
\end{eqnarray}
To get a full (local) coordinate system, one of course has to
choose a third coordinate function, but we leave this choice
arbitrary here. Note that at spatial infinity, $t=V_{12}$ and
$\theta=-3\pi/4$; at any of the $\Gamma_1$-charged centers,
$t=\infty$ and $\theta=0$; and at any of the $\Gamma_2$-charged
centers, $t=\infty$ and $\theta=\pi/2$. Generically, the range of
$t$ on a surface of constant $\theta$ is finite and
$\theta$-dependent. An example (with $N=1$) is shown in fig.\
\ref{tplot}.

We also introduce a $\theta$-dependent ``effective charge''
$\Gamma_\theta$:
\begin{equation}
 \Gamma_\theta \equiv \cos \theta \, \Gamma_1 + \sin \theta
 \, \Gamma_2 \, .
\end{equation}
Then we can rewrite equations (\ref{bps2c1})--(\ref{bps2c3}) on a
surface of fixed $\theta$ as:
\begin{eqnarray} \label{bps2c5}
 2 \, e^{-U} \im[e^{-i \alpha} Z_1] &=& - \langle
 \Gamma_1, \Gamma_\theta \rangle \, t \,, \\
 \label{bps2c6}
2 \, e^{-U} \im[e^{-i \alpha} Z_2] &=& - \langle
 \Gamma_2, \Gamma_\theta \rangle \, t \,, \\
 \label{bps2c7}
2 \, e^{-U} \im[e^{-i \alpha} Z^\perp_L] &=& 2 \im[e^{-i
 \alpha} Z^\perp_L]_{r=\infty} \, ( = \mbox{const.}) \, ,
\end{eqnarray}
or, going back to the compact form of the equations:
\begin{equation} \label{effsimple}
 2 \, \partial_t \left[e^{-U} \im(e^{-i \alpha} \Omega) \right] = -
 \Gamma_\theta.
\end{equation}
This, together with the asymptotics of $\alpha$ at spatial
infinity in (\ref{alphacondition}), implies
\begin{equation} \label{alphaexpr}
 \alpha = \arg Z(\Gamma_\theta) \quad \mbox{or} \quad \alpha =
 \arg[- Z(\Gamma_\theta)] \, .
\end{equation}

Thus, comparing with (\ref{bpscomp}), we see that if $\alpha =
\arg Z(\Gamma_\theta)$, (\ref{effsimple}) describes nothing but
(part of) an ordinary single center flow for a charge
$\Gamma_\theta$, while in the other case, $\alpha = \arg[-
Z(\Gamma_\theta)]$, it describes part of an {\em inverted}
\footnote{An inverted flow is a flow with reversed flow evolution
parameter, here $\partial_t \to -\partial_t$. As noted under
(\ref{bpscomp}), if the flow parameter gets too large, a solution
corresponding to an inverted flow always blows up into a very
unphysical singularity \cite{branessugra}. However, here $t$ is
generically bounded, leaving the possibility to have indeed a well
behaved solution involving (partial) inverted effective subflows.}
single center flow for a charge $\Gamma_\theta$. The only
difference with (\ref{bpscomp}) is the spatial parametrization of
the flow. In the original case we had $\tau=1/r$ going from $0$ to
$\infty$, here we have $t$ as in (\ref{tdef}), which has a
$\theta$-dependent range.

An important question is which of the two possibilities in
(\ref{alphaexpr}) is satisfied at a given point $\sx$. This is
going to to be $\sx$-dependent, since the asymptotic conditions
(\ref{alphacondition}) imply $\alpha = \arg[-Z_\theta]$ at spatial
infinity, and $\alpha = \arg[Z_\theta]$ when approaching any of
the centers. On the other hand, since $\alpha$ and $Z_\theta$ have
to be continuous functions, the spatial surface on which the
solution flips from one possibility to the other must have
$Z_\theta=0$, that is, $Z_2/Z_1=- \tan \theta \in \IR$; in other
words, the moduli are at $(\Gamma_1,\Gamma_2)$- or at
$(\Gamma_1,-\Gamma_2)$-marginal stability, for $\tan \theta \leq
0$ resp.\ $\tan \theta \geq 0$. The converse statement is also
true, provided $t \neq 0$: if the moduli are at $(\Gamma_1,\pm
\Gamma_2)$-marginal stability at a certain point $\sx$ and $t \neq
0$, then $Z_\theta=0$. This follows directly from equations
(\ref{bps2c6})--(\ref{bps2c7}).

Because a surface of marginal stability has real codimension one
in moduli space, and because of the asymptotics of
(\ref{alphacondition}), a surface of $(\Gamma_1,\pm
\Gamma_2)$-marginal stability will in any case split the image of
the solution in moduli space in two parts, as depicted in fig.\
\ref{fattened}: a region connected to the moduli at spatial
infinity, where $alpha=\arg[-Z_\theta]$, and a region connected to
the moduli at the centers, where $\alpha=\arg[Z_\theta]$. The
corresponding regions in space are separated by a surface
$\Sigma_0$ where $Z_\theta=0$ for some value of $\theta$.
Furthermore, we can assume the marginal stability surface under
consideration to be of $(\Gamma_1,\Gamma_2)$ type (so $\tan \theta
\leq 0$ on $\Sigma_0$ if $t \neq 0$), as it is not possible to
have only a $(\Gamma_1,-\Gamma_2)$-marginal stability surface
intersecting the image in moduli space. This will become clear
further on, but can be seen indirectly by imagining to vary
continuously the moduli at spatial infinity to let them approach
the intersecting surface of marginal stability; according to
(\ref{distconstr2}), the configuration will then degenerate to one
with infinitely separated $\Gamma_1$- and $\Gamma_2$-centers, and
the total mass will simply equal the sum of the BPS masses of the
individual charges in the given vacuum. This total mass can only
saturate the BPS bound if the marginal stability surface under
consideration is of $(\Gamma_1,\Gamma_2)$-type.

Note that the solution will not necessarily exist: as noted under
(\ref{bpscomp}), complete inverted flows break down at a finite
value of the flow parameter. Therefore, the $\Gamma_\theta$-flows
for which $\alpha=\arg(-Z_\theta)$ should have a range of $t$ that
is not too large. We will study this problem in more detail below.

\FIGURE[t]{\epsfig{file=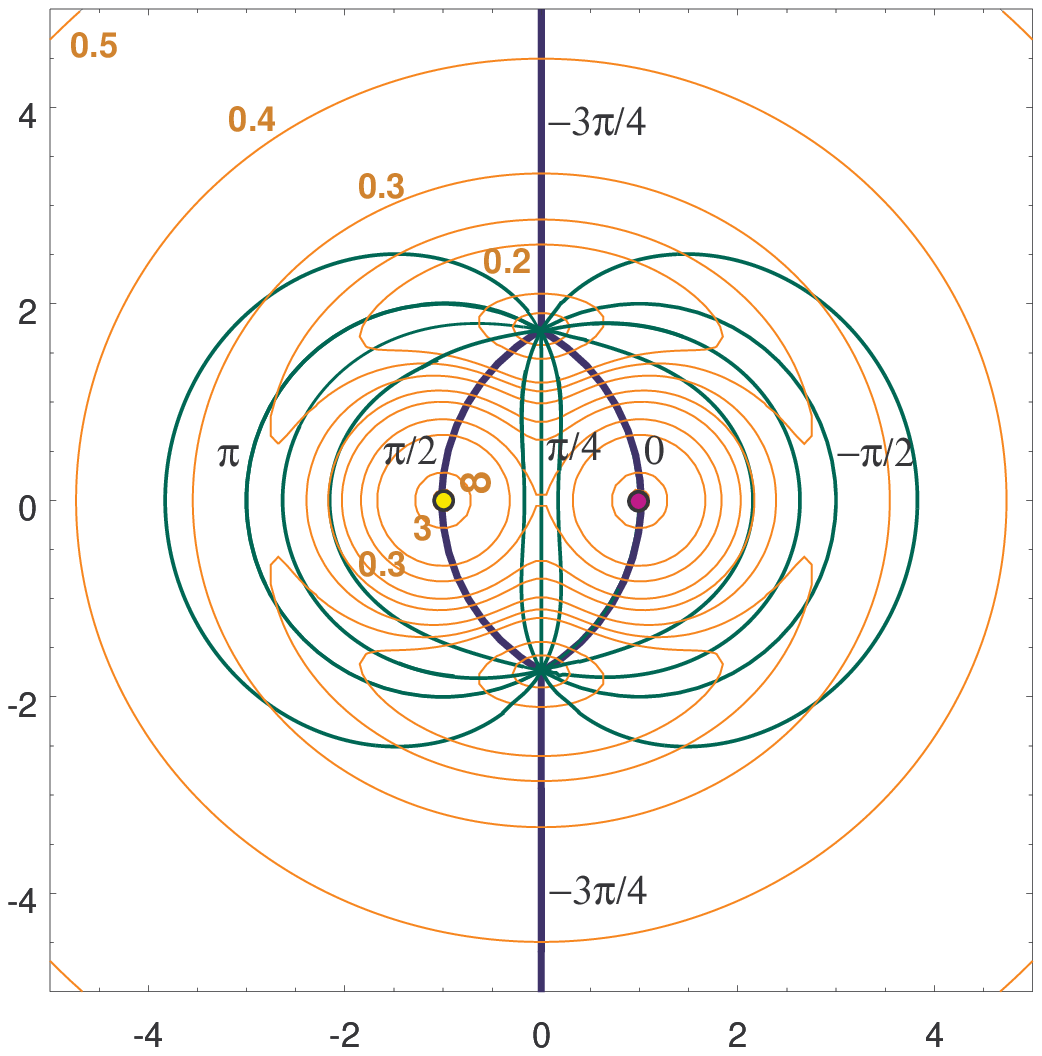,height=9cm}
\caption{Surfaces of equal $t$ (thin orange lines) and equal
$\theta$ (fatter green and even fatter blue lines) for a two
center system with centers at $\sx=(-1,0,0)$ and $\sx=(1,0,0)$.
Equal-$t$ surfaces with $t=0.2, 0.3, 0.4, 0.5, 3, \infty$ are
indicated, as well as equal-$\theta$ surfaces with $\theta=0,
\pi/4, \pi/2, \pi, -3\pi/4, -\pi/2$. The coordinate $t$ is minimal
and equal to zero at the intersection locus of the equal-$\theta$
surfaces, and maximal (=$\infty$) at the positions of the centers.
The fat blue lines ($\theta=0, \pi/2, -3\pi/4$) map to the
skeleton split flow in moduli space.}\label{tplot}}

\FIGURE[t]{\epsfig{file=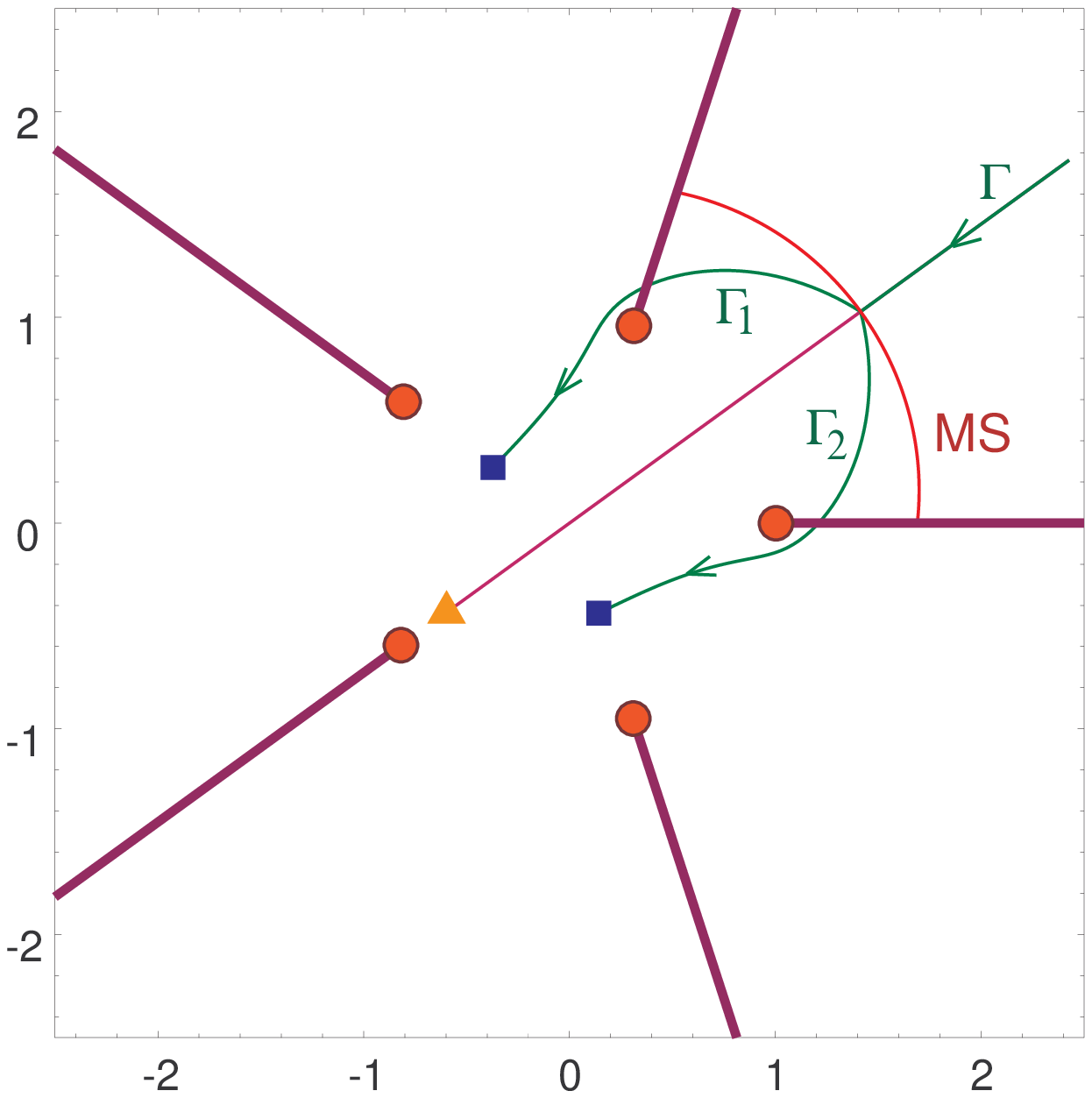,width=0.8\textwidth}
\caption{Numerically computed split flow for the example of fig.
\ref{potential}. A blue square indicates an attractor point with
nonzero $|Z|$, an orange rectangle represents a regular zero. For
graphical purposes, it is convenient to work in the $w$-plane,
where $\arg w \equiv \arg \psi$ and $|w|=\frac{\ln (|\psi|+1)}{\ln
2}$. The horizontal and vertical axis indicates $\re w$ resp.\
$\im w$. The red dots are the five copies of the conifold point in
this 5-fold cover of moduli space. MS is the
$(\Gamma_1,\Gamma_2)$-marginal stability line. The purple line
shows the continuation of the would-be $\Gamma_1+\Gamma_2$-flow
beyond the split point.}\label{split}}

To show how the partition in effective single center flows can be
used to explicitly construct a (possibly numerical) solution, and
thus to establish its existence, let us focus on the $N=1$ case
shown in fig.\ \ref{tplot}. There exists a circle in space where
$t=0$. From this locus, effective (generically partial, and
possibly inverted) $\Gamma_\theta$-flows start for arbitrary
values of $\theta$, together covering all of space. On the surface
$\theta=-3\pi/4$, running to spatial infinity, the effective flow
corresponds to an inverted flow with charge
$\Gamma_\theta=-\frac{1}{\sqrt{2}}(\Gamma_1+\Gamma_2)$, which
image in moduli space is identical to that of a
$\Gamma_1+\Gamma_2$-flow. When $\theta=0$, we have a pure,
complete $\Gamma_1$-flow, and when $\theta=\pi/2$ a pure, complete
$\Gamma_2$-flow. At $t=0$, the moduli must be at
$(\Gamma_1,\Gamma_2)$-marginal stability, with $\alpha=\arg
Z_1=\arg Z_2$ (this follows directly from
(\ref{bps2c6})-(\ref{bps2c7}) plus the asymptotic conditions
(\ref{alphacondition})). Hence this point in moduli space is
determined as the intersection of the $\Gamma_1+\Gamma_2$ flow
starting from the moduli at spatial infinity with the surface of
marginal stability. The $\theta=-3\pi/4$, $\theta=0$ and
$\theta=\pi/2$ flows together form a ``split flow'', as shown for
a specific (numerically computed) quintic example in fig.\
\ref{split}. The (partial) flows for the other values of $\theta$
will fatten this split flow to something like fig.\
\ref{fatconstr} (b). A subset of the flows with $\tan \theta < 0$
will cross a zero of $Z_\theta$, namely where the surface of
marginal stability is crossed in moduli space, as explained
earlier and illustrated in fig.\ \ref{fatconstr}. Note that this
does not lead to a breakdown of the solution, since at the same
time, we jump from $\alpha=\arg(-Z_\theta)$ to $\alpha=\arg
Z_\theta$, such that the inverted $\Gamma_\theta$-flow we have up
to the zero gets smoothly connected to an ordinary
$\Gamma_\theta$-flow starting from the zero.

\FIGURE[t]{\epsfig{file=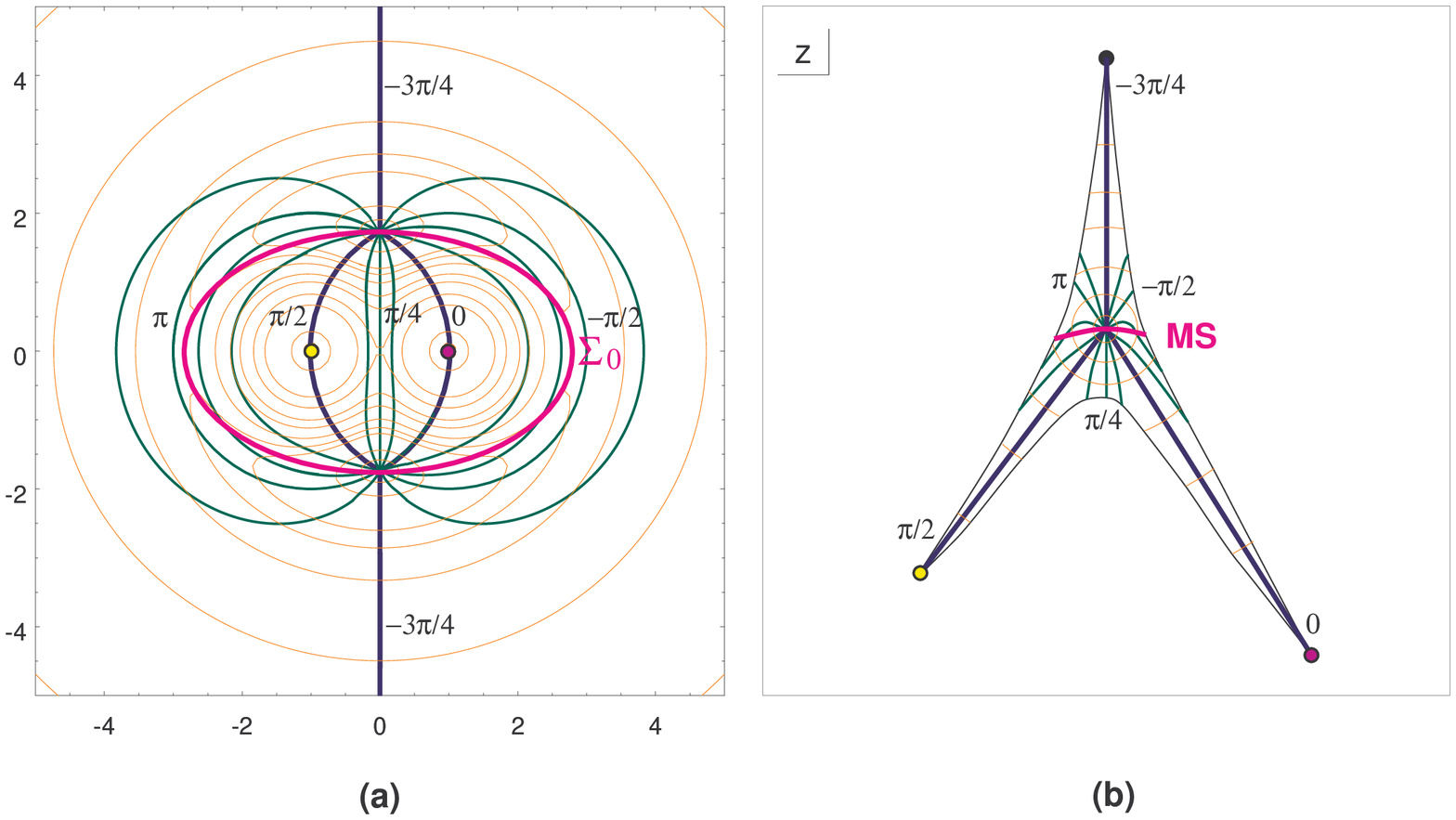,width=1.1\textwidth}
\caption{(a): Same as fig.\ \ref{tplot}, with additionally a
possible location of the surface $\Sigma_0$ sketched (where the
$(\Gamma_1,\Gamma_2)$-marginal stability line is crossed and the
condition $\alpha=\arg(\pm Z_\theta)$ flips). (b): Sketch of a
possible corresponding image in moduli space, including the image
of some lines of constant $\theta$ and $t$, and of $\Sigma_0$
(indicated as MS).} \label{fatconstr}}

This construction shows that a multicenter solution for given
center locations satisfying the constraint (\ref{distconstr2}),
will indeed exist, if each of the $\Gamma_\theta$ flows exists.
The latter will be the case provided none of the $\Gamma_\theta$
flows crosses the $(\Gamma_1,-\Gamma_2)$-marginal stability
surface (where $\tan \theta \geq 0$, $Z_\theta$ vanishes and the
$\alpha=\arg( \pm Z_\theta)$ condition flips), or, if some do,
provided their $t$ range is not too large.\footnote{The solution
to the second BPS equation (\ref{mc2}) does not present further
obstacles to the existence of the solution, and is discussed
below.} It is quite plausible that this will be satisfied if and
only if the ``skeleton'' split flow exists, though we will not try
to prove this here. What could go wrong for example is that the
partial $\Gamma_\theta$-flow for $\theta=\pi/4$, which finitely
extends the incoming $\Gamma_1+\Gamma_2$-branch of the split flow,
could hit a regular zero and have a maximal $t$ beyond the point
where the inverted flow beyond the zero blows up. A necessary
condition for this to happen is of course that the complete
$\Gamma_1+\Gamma_2$-flow hits a regular zero. Such cases (as in
the example of fig.\ \ref{split}) are quite interesting on their
own, as they correspond to states that can only be realized as a
multicenter solution; in particular they {\em cannot} be realized
as a regular black hole.\footnote{It is not that easy to find such
examples with regular black holes as constituents, because if the
charges $\Gamma_1 \, (=\Gamma_{\theta=0})$ and $\Gamma_2 \,
(=\Gamma_{\theta=\pi/2})$ flow to a nonzero minimal $|Z|$, then
$\Gamma_1+\Gamma_2 \, (= \sqrt{2} \, \Gamma_{\theta=\pi/4})$
usually flows to a finite minimum as well. One basically needs an
obstruction for smooth $\theta: 0 \to \pi/2$ interpolation between
the $\Gamma_1$ and $\Gamma_2$ attractor flows. In the example of
fig.\ \ref{split}, the obstruction occurs due to the conifold
points ``inside'' the split. This observation presumably also
explains the apparent absence of such examples for e.g. the one
modulus $T^6$ compactification of \cite{M}.}

Existence of the full multicenter solution will certainly be
implied by existence of the skeleton split flow for many-center
configurations approximating the idealized, uniformly charged
spherical shell of fig.\ \ref{shells}. Indeed, when we uniformly
distribute an enormous number $N$ of $\Gamma_2$- centers on a
sphere at equilibrium distance $r=r_{\mathrm{ms}}$ from a black
hole center with charge $N \Gamma_1$, the corresponding fattened
flow given by the multicenter solution will in fact be very thin,
staying everywhere very close to the skeleton split flow. In the
limit $N \to \infty$, the fattened flow becomes infinitely thin
and reduces to the split flow itself. The $\Gamma_1+\Gamma_2$
branch corresponds to the solution outside\footnote{The shell will
have a radius $r_{\mathrm{ms}}$ proportional to $N$ in the
$\sx$-coordinates. The typical distance $\ell$ between centers on
the sphere is of order $\sqrt{N}$. So the discrete structure of
the charge becomes visible at a distance of order $\ell \sim
r_{\mathrm{ms}}/\sqrt{N}$ from the shell. By ``outside'' or
``inside'' the shell, we mean being at a much greater distance
from it than $\ell$, such that the discrete structure is
essentially invisible.} the $\Gamma_1$ branch to the solution
inside the shell, and the $\Gamma_2$ branch to the solution on
(and near) the shell. This can be deduced directly from the above
equations. It is plausible because away from the shell, in the
limit $N \to \infty$ and on the scale of $r_{\mathrm{ms}}$, the
situation reduces effectively to the idealized one depicted in
fig. \ref{shells}. On the other hand, when zooming in to the
natural $\sx$-coordinate scale close to the shell, one sees a
number of $\Gamma_2$ centers, with separations of order $\sqrt{N}
\to \infty $, floating around in a background with moduli value at
$z^a_{\mathrm{ms}}$. Around each of those centers, we should
therefore have a moduli flow corresponding to the
$\Gamma_2$-branch of the split flow.

Actually, to get the full solution, we should also construct
$\omega$ from (\ref{bps2c4}). We did not do this explicitly, but
because the position constraint (\ref{distconstr2}) is in fact the
integrability condition for this equation, a solution should
certainly exist. In terms of $t$ and $\theta$, (\ref{bps2c4})
reads, rather nicely:
\begin{equation} \label{omegatheta}
\s* d \, \omega = - \langle \Gamma_1,\Gamma_2 \rangle \, t^2 \,
d\theta \, .
\end{equation}
Here $\s* d \, \omega$ is actually the invariant local quantity
(because $\omega$ can always be transformed to any $\omega +
d\lambda$ by an $\sx$-dependent time coordinate transformation),
and hence of more physical significance than $\omega$ itself.
Equation (\ref{omegatheta}) is quite useful to visualize this
quantity in specific configurations.

So our conclusion is: {\em (at least some) regular multicenter BPS
black hole solutions exist if and only if the corresponding split
flow solution exists.}

A final remark we want to make is that for multicenter BPS
solutions involving also empty holes, this conclusion apparently
does not hold: even when the split flow exist (with at least one
branch, say $\theta=0$, ending on e.g.\ a conifold point) --- and
consequently also an idealized spherical shell solution --- a
corresponding multicenter solution (with point sources) does not
seem to exist, because any $\Gamma_\theta$-flow with $\theta$
sufficiently small but nonzero, will hit a regular zero, and has
at the same time an arbitrarily large maximal $t$, leading to a
breakdown of the solution. This might be related to the
delocalized nature of the source for empty holes, as discussed in
section \ref{ben}, but at this point, we do not have a concrete
proposal for a resolution of this puzzle.

\section{Composite configurations and existence of BPS states in string theory}
\label{existence}

\subsection{A modified correspondence conjecture}

\FIGURE[t]{\epsfig{file=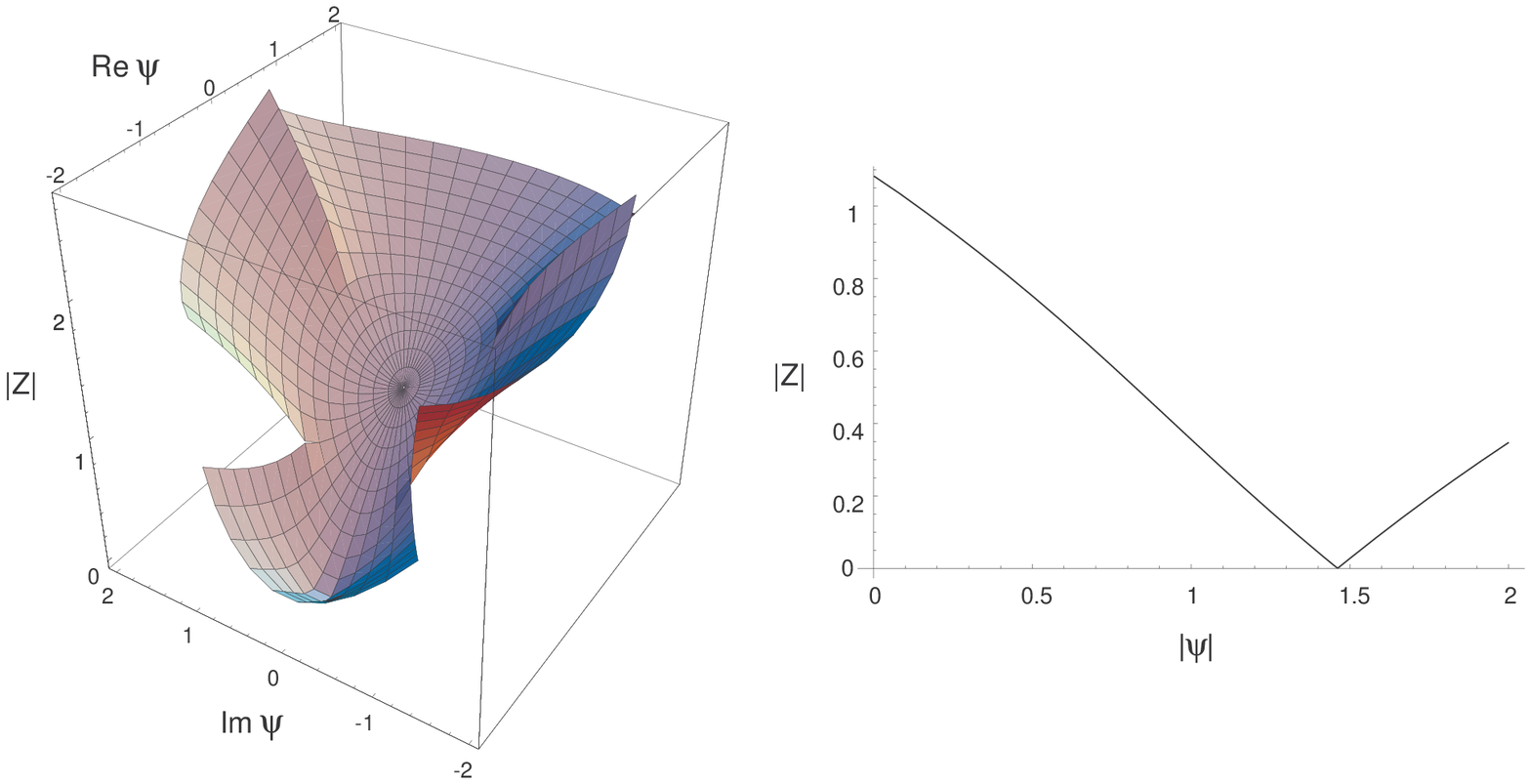,height=7cm} \caption{An
example of a BPS state of type IIA string theory compactified on
the quintic at the Gepner point, which does not have a
corresponding single center black hole description. \emph{Left:}
modulus of the central charge as a function of the quintic modulus
$\psi$ for the BPS state $|10000\rangle_B$ of \cite{BDLR} (with
charge $(Q_6,Q_4,Q_2,Q_0)=(2,0,5,0)$), numerically computed using
mirror symmetry. There is a regular zero at $\psi \approx -1.46$,
hit by the flow starting at the Gepner point $\psi=0$.
\emph{Right:} $|Z|$ as a function of $|\psi|$ on the negative
real~axis.}\label{10000}}

As already mentioned in the introduction, it turns out that there
are quite a few examples of BPS states known to exist in certain
Calabi-Yau compactifications of type II string theory, which do
not have a corresponding single center BPS black (or empty) hole
solution, not even for large $N$. An example is given in fig.\
\ref{10000}. Other examples are the higher dyons and the W-boson
in Seiberg-Witten theory.

Strictly speaking, this disproves the correspondence conjecture of
\cite{M}. In \cite{branessugra}, this puzzle and related paradoxes
were studied in detail, and the necessity of considering more
general stationary (multicenter) solutions in this context was
demonstrated. Thus we are brought to the following adaptation of
the correspondence conjecture, in its strongest form: {\em a BPS
state of a given charge exists in the full string theory if and
only if a single or (possibly multi-) split attractor flow
corresponding to that charge exists.}\footnote{In fact, this
statement of the conjecture is probably too strong, as it might
happen that a certain split flow exists but ceases to do so after
continuous variation of the moduli, {\em without} actually
crossing a surface of marginal stability. In that case, one does
not expect the original state to exist as a BPS state in the full
quantum theory. This is similar to a phenomenon occurring in the
context of 3-pronged strings \cite{threeprong}, where the
existence criterion for certain BPS states needs to be refined
accordingly. We will discuss this issue in more detail in
\cite{quintic}. \label{subtlety}} Note that often, a BPS state can
have several different realizations in the four dimensional low
energy effective supergravity theory, either as an ordinary BPS
black hole, corresponding to a single flow, or as one or more
multicenter solutions (or spherical shell solutions),
corresponding to one or more split flows. In fig.\ \ref{compfl},
it is shown how this modification of the conjecture indeed
resolves the paradoxes as presented by the examples mentioned in
the previous paragraph.

\FIGURE[t]{\epsfig{file=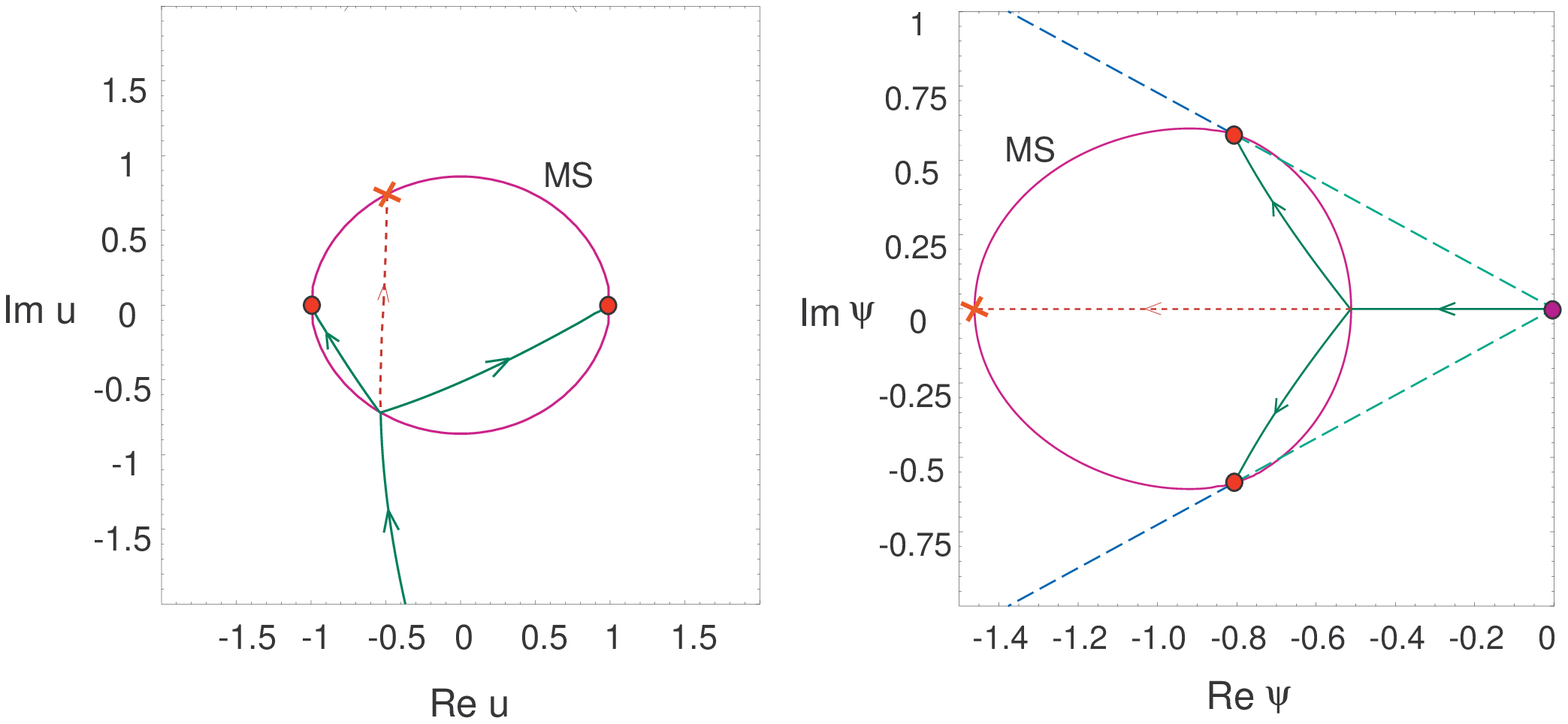,height=6.5cm}
\caption{Numerically computed split flow for a $(2,-1)$-dyon at
weak coupling in Seiberg-Witten theory (left) and for the state
$|10000\rangle_B$ at the Gepner point $\psi=0$ of type IIA on the
quintic (right). The former has $(\Gamma_1,\Gamma_2)=(2
(1,-1),(0,1))$, the latter
$(\Gamma_1,\Gamma_2)=((6,3,19,-10),(-4,-3,-14,10))$. The purple
ellipsoid line is the relevant line of marginal stability, the
dotted red line is the naive $(2,0,5,0)$ attractor flow, crashing
on the regular zero of $Z$ indicated by a red cross. The wedge $4
\pi/5 < \arg \psi < 6 \pi/5$ is indicated by dashed lines. Both
examples happen to involve empty hole charges only.}
\label{compfl}}

A much more detailed study of the BPS spectrum of the quintic from
this effective field theory point of view will be presented in
\cite{quintic}.

\subsection{Marginal stability, Joyce transitions and
$\Pi$-stability} \label{pistab}

From~(\ref{rmsformula}) or (\ref{distconstr}) or
(\ref{distconstr2}), it follows that when the moduli at infinity
approach the the surface of $(\Gamma_1,\Gamma_2)$-marginal
stability, the equilibrium distance between the $\Gamma_1$- and
$\Gamma_2$-sources will diverge, eventually reaching infinity at
marginal stability. Beyond the surface, the realization of this
charge as a BPS $(\Gamma_1,\Gamma_2)$-composite no longer exists.
This gives a nicely continuous four dimensional spacetime picture
for the decay of the state when crossing a surface of marginal
stability.

Furthermore, these formulae tell us at which side of the marginal
stability surface the composite state can actually exist: since
$r_{\mathrm{ms}}>0$, it is the side satisfying
\begin{equation} \label{stabcond}
\langle \Gamma_1,\Gamma_2 \rangle \, \sin(\alpha_1-\alpha_2) > 0
\,,
\end{equation}
where $\alpha_i = \arg Z(\Gamma_i)_{r=\infty}$. Sufficiently close
to marginal stability, this reduces to
\begin{equation}
\langle \Gamma_1,\Gamma_2 \rangle \, (\alpha_1-\alpha_2) > 0\,,
\end{equation}
which is precisely the stability condition for ``bound states'' of
special lagrangian 3-cycles found in a purely Calabi-Yau
geometrical context by Joyce! (under more specific conditions,
which we will not give here)~\cite{joyce,KM}.

Note also that, since the right-hand side of~(\ref{imZ1}) can only
vanish for one value of $\tau$, the composite configurations we
are considering here will actually satisfy
\begin{equation}
|\alpha_1-\alpha_2|<2 \pi \,.
\end{equation}

If the constituent $\Gamma_i$ of the composite configuration for
which $\langle \Gamma,\Gamma_i \rangle > 0$ can be identified with
a ``subobject'' of the state as defined in~\cite{DFR}, the above
conditions imply that the phases satisfy the $\Pi$-stability
criterion introduced in that reference. Though this similarity is
interesting, it is not clear how far it extends. $\Pi$-stability
seems to be considerably more subtle than what emerges here. It
would be interesting to explore this connection further.

\subsection{Non-BPS composites}

\FIGURE[t]{\centerline{\epsfig{file=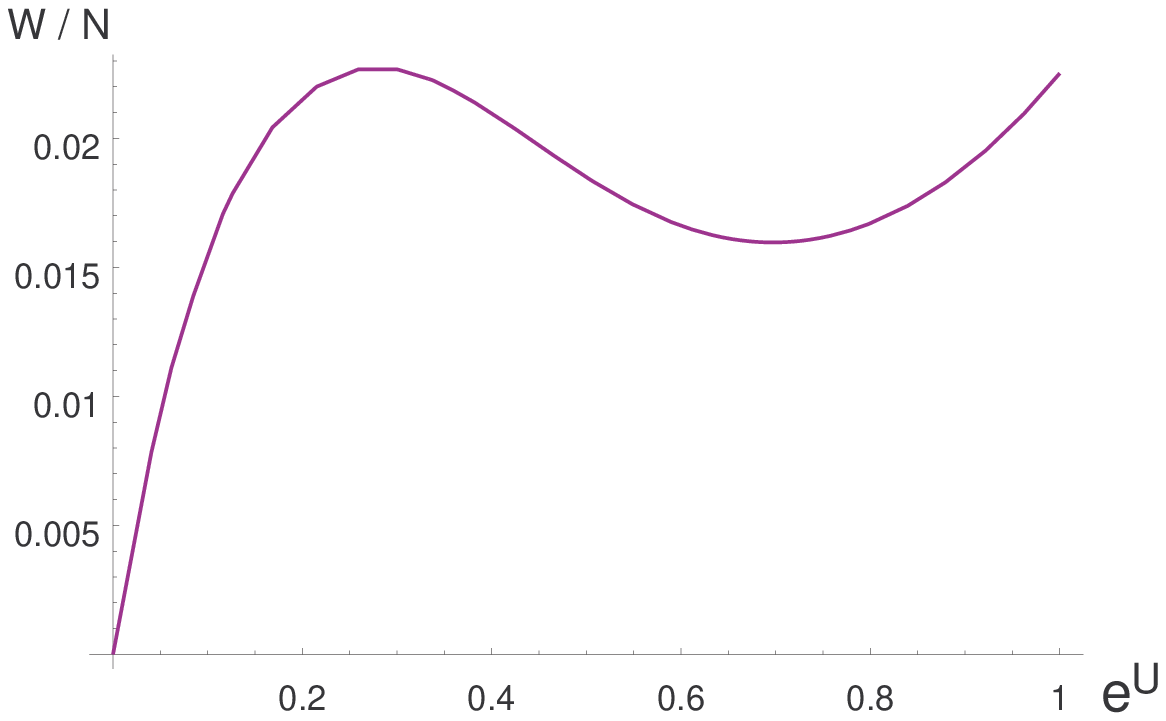,height=6cm}}
\caption{Potential $W$ (as a function of $e^U$) for a test
particle of charge $(Q_6,Q_4,Q_2,Q_0)=(0,0,1,0)$ in the field of a
charge $N (1,-1,-4,1)$ BPS black hole, for a quintic
compactification at $w=10 e^{\pi/5}$ (see fig.\ \ref{split} for
the definition of the modulus $w$). The nonzero minimum indicates
the existence of a classically stable non-BPS configuration. In
this case, a BPS black hole with the same total charge also
exists, so the non-BPS state is possibly quantum mechanically
unstable through tunneling.
 }\label{nonBPSpotential}}

From the discussion of section \ref{stateq}, one could also
contemplate the existence of non-BPS composites. For example, what
happens when we try to throw up to $N$ particles of charge
$\Gamma_2$ into a black hole of charge $N \Gamma_1$, if we know
that a composite BPS $(N \Gamma_1,N \Gamma_2)$-configuration does
not exist? What does the ground state of this system look like in
the low energy effective supergravity theory? One possibility is
that a stable, statinary, non-BPS composite develops. Another
possibility is that at a certain point, we simply cannot throw any
$\Gamma_1$-particle into the black hole anymore, because it is
repelled from it all the way up to spatial infinity. As a first
approach to study non-BPS composites, one can look for nonzero
minima of the force potential $W$ for a test particle in the field
of a (BPS) black hole, equation (\ref{potexpr}). A configuration
with this particle in its equilibrium position can then be
expected to exist as a non-BPS solution.

A quintic example illustrating the possibility of having such a
nonzero minimum is given in fig.\ \ref{nonBPSpotential}.

\section{Conclusions}
\label{conclusions}

We have proposed a modified version of the correspondence
conjecture of \cite{M} between BPS states in Calabi-Yau
compactifications of type II string theory on the one hand, and
four dimensional stationary ${\cal N}=2$ supergravity solutions on
the other hand, and established a link between these solutions and
split attractor flows in moduli space. Some interesting
connections emerged, to the enhan\c{c}on mechanism, the 3-pronged
string picture of QFT BPS states, $\Pi$-stability and Joyce
transitions of special lagrangian manifolds.

The most prominent open question is of course whether the
conjecture (perhaps in a more refined form, as outlined in
footnote \ref{subtlety}) actually works. A more systematic
comparison with known string theory results would be the obvious
strategy for verifying this, but this is complicated considerably
because of the fact that the number of cases that are accessible
in both approaches at the same time, is rather limited. Some steps
in that direction will be presented in \cite{quintic}.

Other interesting open question are: How to find a proper
``multicenter'' description of composite configurations involving
empty holes? Is there a connection between D-brane moduli spaces
and supergravity solution moduli spaces? Can those solution moduli
spaces teach us something about the entropy of these states? What
is the precise relation with $\Pi$-stability? Is there a link
between the emergence of spatially extended configurations here
and in the context of noncommutative brane effects, as for example
in \cite{myers}? And who's going to win the Subway Series, the
Yankees or the Mets?

Hopefully some of these question will get an answer in the near
future.

\acknowledgments

I would like to thank Mike Douglas, Tomeu Fiol, Brian Greene, Greg
Moore, Mark Raugas and Christian R\"omelsberger for useful
discussions and correspondence, and the conference organizers Eric
D'Hoker, D.H.\ Phong and S.T.\ Yau for their hard work and
patience. Part of this work was done in collaboration with Brian
Greene and Mark Raugas.

\newcommand{\dgga}[1]{\href{http://xxx.lanl.gov/abs/dg-ga/#1}{\tt
dg-ga/#1}}
\renewcommand\baselinestretch{1.08}

\normalsize

\begin{thebibliography}{99}

\bibitem{SW}
N.~Seiberg and E.~Witten, \emph{Electric-magnetic duality,
monopole condensation and confinement in $N=2$ supersymmetric
Yang-Mills theory}, \npb{426}{1994}{19} [\hepth{9407087}].

\bibitem{joyce}
D.~Joyce, \emph{On counting special lagrangian homology
3-spheres}, \hepth{9907013}.

\bibitem{hitchin}
N. Hitchin, \emph{The moduli space of special lagrangian
submanifolds}, \dgga{9711002}.

\bibitem{RS}
A.~Recknagel and V.~Schomerus, \emph{D-branes in Gepner models},
\npb{531}{1998}{185} [\hepth{9712186}].

\bibitem{BDLR}
I.~Brunner, M.R. Douglas, A.~Lawrence and C.~R\"omelsberger,
  \emph{D-branes on the quintic}, \jhep{08}{2000}{015}
  [\hepth{9906200}].


\bibitem{D}
M.R. Douglas, \emph{Topics in D-geometry}, \cqg{17}{2000}{1057}
[\hepth{9910170}].

\bibitem{DR}
D.-E. Diaconescu and C.~Romelsberger, \emph{D-branes and bundles
on
  elliptic fibrations}, \npb{574}{2000}{245} [\hepth{9910172}].

\bibitem{Sch}
E.~Scheidegger, \emph{D-branes on some one- and two-parameter
  Calabi-Yau hypersurfaces}, \jhep{04}{2000}{003} [\hepth{9912188}].

\bibitem{BS}
I.~Brunner and V.~Schomerus, \emph{D-branes at singular curves of
  Calabi-Yau compactifications}, \jhep{04}{2000}{020}
  [\hepth{0001132}].

\bibitem{DFR}
M.R. Douglas, B.~Fiol and C.~Romelsberger, \emph{Stability and BPS
  branes}, \hepth{0002037}.

\bibitem{DFR2}
M.R. Douglas, B.~Fiol and C.~Romelsberger, \emph{The spectrum of
BPS
  branes on a noncompact Calabi-Yau}, \hepth{0003263}.

\bibitem{DD}
D.E. Diaconescu and M.R. Douglas, \emph{D-branes on Stringy
Calabi-Yau Manifolds}, \hepth{0006224}.

\bibitem{FM}
B.~Fiol and M.~Marino, \emph{BPS states and algebras from
quivers}, \jhep{07}{2000}{031} [\hepth{0006189}].

\bibitem{KM}
S.~Kachru and J.~McGreevy, \emph{Supersymmetric three-cycles and
  (super)symmetry breaking}, \prd{61}{2000}{026001} [\hepth{9908135}].

\bibitem{FKS}
S.~Ferrara, R.~Kallosh and A.~Strominger, \emph{$N=2$ extremal
black
  holes}, \prd{52}{1995}{5412} [\hepth{9508072}].

\bibitem{M}
G.~Moore, \emph{Arithmetic and attractors}, \hepth{9807087};
\emph{Attractors and arithmetic}, \hepth{9807056}.

\bibitem{branessugra}
F.~Denef, \emph{Supergravity flows and D-brane stability},
\jhep{08}{2000}{050} [\hepth{0005049}].

\bibitem{enhancon}
C.V. Johnson, A.W. Peet and J.~Polchinski, \emph{Gauge theory and
the
  excision of repulson singularities}, \prd{61}{2000}{086001}
  [\hepth{9911161}].

\bibitem{BLS}
K.~Behrndt, D.~L\"ust and W.A. Sabra, \emph{Stationary solutions
of
  $N=2$ supergravity}, \npb{510}{1998}{264} [\hepth{9705169}].

\bibitem{CWKM}
G.L. Cardoso, B.~de Wit, J.~Käppeli and T.~Mohaupt,
\emph{Stationary BPS Solutions in N=2 Supergravity with
$R^2$-Interactions}, \hepth{0009234}.

\bibitem{quintic}
F.~Denef, B.~Greene, M.~Raugas, \emph{Type IIA D-branes on the
Quintic from a four dimensional supergravity perspective}, to
appear.

\bibitem{SG}
B.~de~Wit, P.G. Lauwers and A.V. Proeyen, \emph{Lagrangians of
$N=2$
  supergravity-matter systems}, \npb{255}{1985}{569};\\
B.~Craps, F.~Roose, W.~Troost and A.V. Proeyen, \emph{What is
special
  kaehler geometry?}, \npb{503}{1997}{565} [\hepth{9703082}].

\bibitem{FGK}
S.~Ferrara, G.W. Gibbons and R.~Kallosh, \emph{Black holes and
  critical points in moduli space}, \npb{500}{1997}{75}
  [\hepth{9702103}].

\bibitem{BPS37}
A.~Sen, \emph{BPS states on a three brane probe},
\prd{55}{1997}{2501} [\hepth{9608005}].

\bibitem{threeprong}
M.R. Gaberdiel, T.~Hauer and B.~Zwiebach, \emph{Open string-string
  junction transitions}, \npb{525}{1998}{117} [\hepth{9801205}];\\
O.~Bergman and A.~Fayyazuddin, \emph{String junctions and BPS
states
  in seiberg-witten theory}, \npb{531}{1998}{108} [\hepth{9802033}];\\
A.~Mikhailov, N.~Nekrasov and S.~Sethi, \emph{Geometric
realizations
  of BPS states in $N=2$ theories}, \npb{531}{1998}{345}
  [\hepth{9803142}];\\
O.~DeWolfe, T.~Hauer, A.~Iqbal and B.~Zwiebach, \emph{Constraints
on
  the BPS spectrum of $N=2$, $D = 4$ theories with ADE flavor
  symmetry}, \npb{534}{1998}{261} [\hepth{9805220}].

\bibitem{S}
A.~Strominger, \emph{Massless black holes and conifolds in string
theory}, \npb{451}{1995}{96} [\hepth{9504090}].

\bibitem{modulispace}
R. Ferell and D. Eardley, \emph{Slow-motion scattering and
coalescence of maximally charged black holes}, \emph{Phys.\ Rev.\
}{\bf 59} (1987) 1617.

\bibitem{adsfrag}
J.~Maldacena, J.~Michelson and A.~Strominger, \emph{Anti-de~Sitter
  fragmentation}, \jhep{02}{1999}{011} [\hepth{9812073}].

\bibitem{AdSCFT}
J.~Maldacena, \emph{The large-$N$ limit of superconformal field
  theories and supergravity}, \atmp{2}{1998}{231} [\hepth{9711200}];\\
S.S. Gubser, I.R. Klebanov and A.M. Polyakov, \emph{Gauge theory
  correlators from non-critical string theory}, \plb{428}{1998}{105}
  [\hepth{9802109}];\\
E.~Witten, \emph{Anti-de~Sitter space and holography},
\atmp{2}{1998}{253} [\hepth{9802150}].

\bibitem{GLaz}
B.R. Greene and C.I. Lazaroiu, \emph{Collapsing D-branes in
Calabi-Yau
  moduli space, 1}, \hepth{0001025}.

\bibitem{MTW}
C. Misner, K. Thorne and J.A. Wheeler, \emph{Gravitation}, Freeman
and Co.\ 1973, chapter~21.

\bibitem{myers}
R.C. Myers, \emph{Dielectric-branes}, \jhep{12}{1999}{022}
[\hepth{9910053}];\\ N.R. Constable, R.C. Myers and O.~Tafjord,
\emph{The noncommutative
  bion core}, \prd{61}{2000}{106009} [\hepth{9911136}].


\end{thebibliography}
\end{document}